\documentclass[a4paper,fleqn,usenatbib]{mnras}

\usepackage{newtxtext,newtxmath}

\usepackage[T1]{fontenc}

\DeclareRobustCommand{\VAN}[3]{#2}
\let\VANthebibliography\thebibliography
\def\thebibliography{\DeclareRobustCommand{\VAN}[3]{##3}\VANthebibliography}

\usepackage{graphicx}	
\usepackage{amsmath}	
\usepackage{pdflscape}
\usepackage{ae,aecompl}
\usepackage{float}
\usepackage{tabularx}

\title[\textit{Gaia} orbits for UV excess binaries]{The white dwarf binary pathways survey - X. \textit{Gaia} orbits for known UV excess binaries.}

\author[J. A. Garbutt et al.]{
J. A. Garbutt,$^{1}$\thanks{E-mail: jagarbutt1@sheffield.ac.uk}
S. G. Parsons,$^{1}$
O. Toloza,$^{2,3}$
B. T. Gänsicke,$^{4}$
M. S. Hernandez,$^{2,3}$
\newauthor
D. Koester,$^{5}$
F. Lagos,$^{4}$
R. Raddi,$^{6}$
A. Rebassa-Mansergas,$^{6,7}$
J. J. Ren,$^{8}$
\newauthor
M. R. Schreiber,$^{2,3}$
M. Zorotovic$^{9}$
\\
$^{1}$Department of Physics \& Astronomy, University of Sheffield, Sheffield S3 7RH, UK\\
$^{2}$Departamento de F{\'i}sica, Universidad T{\'e}cnica Federico Santa Mar{\'i}a, Avenida España 1680, Valpara{\'i}so, Chile\\
$^{3}$Millennium Nucleus for Planet Formation, NPF, Valparaıso, Av. España 1680, Chile\\
$^{4}$Department of Physics, University of Warwick, Coventry CV4 7AL, UK\\
$^{5}$Institut für Theoretische Physik und Astrophysik, University of Kiel, 24098 Kiel, Germany\\
$^{6}$Departament de Física, Universitat Politècnica de Catalunya, c/Esteve Terrades 5, E-08860 Castelldefels, Spain\\
$^{7}$Institute for Space Studies of Catalonia, c/Gran Capità 2-4, Edif. Nexus 201, E-08034 Barcelona, Spain\\
$^{8}$Key Laboratory of Space Astronomy and Technology, National Astronomical Observatories, Chinese Academy of Sciences, Beĳing 100101, P. R. China\\
$^{9}$Instituto de Física y Astronomía, Universidad de Valparaíso, Av. Gran Bretaña 1111, Valparaíso, Chile
}

\date{Accepted XXX. Received YYY; in original form ZZZ}

\pubyear{2023}

\begin{document}
\label{firstpage}
\pagerange{\pageref{firstpage}--\pageref{lastpage}}
\maketitle

\begin{abstract}
\noindent
White dwarfs with a F, G or K type companion represent the last common ancestor for a plethora of exotic systems throughout the galaxy, though to this point very few of them have been fully characterised in terms of orbital period and component masses, despite the fact several thousand have been identified. \textit{Gaia} data release 3 has examined many hundreds of thousands of systems, and as such we can use this, in conjunction with our previous UV excess catalogues, to perform spectral energy distribution fitting in order to obtain a sample of 206 binaries likely to contain a white dwarf, complete with orbital periods, and either a direct measurement of the component masses for astrometric systems, or a lower limit on the component masses for spectroscopic systems. Of this sample of 206, four have previously been observed with {\it Hubble Space Telescope} spectroscopically in the ultraviolet, which has confirmed the presence of a white dwarf, and we find excellent agreement between the dynamical and spectroscopic masses of the white dwarfs in these systems. We find that white dwarf plus F, G or K binaries can have a wide range of orbital periods, from less than a day to many hundreds of days. A large number of our systems are likely post-stable mass transfer systems based on their mass/period relationships, while others are difficult to explain either via stable mass transfer or standard common envelope evolution.
\end{abstract}




\section{Introduction}
\label{sec:Introduction}

White Dwarfs (hereafter, WDs) are the last stage of the evolution of low to intermediate mass stars, of initial mass around $\sim 0.85 - 9$\,M\textsubscript{$\odot$} \citep{Kepler_2007,Cummings_2018}. It has been found through observations (e.g. \citealt{Holberg_2009, Toonen_2017}), that around 18-26 per cent of WDs are in a binary system, meaning that the WD evolved in proximity to a companion. 

If the initial binary system had an orbital period, $P$, of less than roughly $10^4$ days, then it is likely that the two stars interacted as the WD progenitor evolved off the main-sequence \citep{Willems_2004}. A possible interaction would be a common envelope phase \citep{Paczynski_1976}. Here, as the more massive star's envelope expands and overflows its Roche lobe, it begins unstably shedding matter. In this process, the less massive star also overfills its Roche lobe, resulting in the formation of a common envelope around the core of the donor and the less massive companion. This causes them to spiral inward. If enough orbital energy is transferred to the envelope to unbind and eject it, a post-common envelope system with a shorter period is formed \citep{Rebassa-Mansergas_2008, Nebot_2011}. 

The specific dynamics of the common envelope phase are difficult to reproduce with hydrodynamical modelling \citep{Passy_2012, Ohlmann_2016, Ondratschek_2022, Moreno_2022}. As such, typically a simplified equation involving the common envelope efficiency, $\alpha_\mathrm{CE}$, is used instead - where $\alpha_\mathrm{CE}$ is the fraction of the change in orbital energy used to unbind the envelope. Therefore, a lower efficiency implies a greater reduction in the orbital period. \citet{Zorotovic_2010} has found a value of $\alpha_\mathrm{CE} \approx 1/3$ through observations of WD\,+\,dM binaries, with similarly low efficiencies being found for WD\,+\,BD binaries \citep{Zorotovic_2022} and close low mass WD\,+\,WD binaries \citep{Scherbak_2023} - though it is uncertain if such a value is universal. If this is the case, then most WD\,+\,FGK systems would be expected to emerge from the common envelope phase with periods too short for them to survive a second common envelope phase, meaning that forming double-degenerate systems via this pathway would be extremely challenging. Indeed, it was this issue that lead to the creation of the so-called $\gamma$ formalism \citep{Nelemans_2000}. Gamma formalism, also called ``common envelope without spiral in", is stable but highly non-conservative mass transfer. However, there are alternative channels to the creation of longer period WD\,+\,FGK binaries that do not involve a common envelope phase, such as the stable mass transfer channel \citep{Webbink_2008}.

Stable mass transfer, as proposed by \citet{Webbink_2008}, is thought to occur when the masses of the two stars are very similar, or if the system comes into contact when the donor star is on, or has just left, the main sequence. In this scenario, the less massive star is likely to successfully accrete the mass overflowing from the more massive companion at a steady rate, which can lead to a widening of the binary if the transfer is non-conservative \citep{Podsiadlowski_2014}. This would leave the system with a wide enough period to survive a common envelope phase without merging when the lower mass star evolves, which can lead to double-degenerate systems, a progenitor for thermonuclear supernovae. It is not yet understood how conservative this mechanism is in practice, with \citet{Kawahara_2018} suggesting the mass transfer is usually non-conservative, whilst \citet{Podsiadlowski_2014} suggests a near fully conservative transfer.

Until recently, the majority of WD\,+\,FGK binaries with known orbital periods (e.g. \citealt{Parsons_2015, PathwayIV, PathwayVI, PathwayVII, PathwayVIII}) were short period systems of around $\sim\,0.5 - 2.5$\,days, which can be reproduced with the same common envelope efficiency as used for WD\,+\,dM binaries ($\alpha_\mathrm{CE} \approx 1/3$). These short period systems can go on to become cataclysmic variables or supersoft X-ray source systems. Longer period systems had proven more elusive, though there were a small number published \citep{KruseAgol_2014, Kawahara_2018, Masuda_2019, PathwayIX, Yamaguchi_2023}, with \citet{Shahaf_2023b} identifying a few thousand WD\,+\,FGK candidates in \textit{Gaia} data release 3, with periods around $\approx 100 - 1000$ days - though this sample has yet to be explored in great detail. These longer period systems are likely the progenitors of double degenerate systems and symbiotic binaries \citep{Zorotovic_2014}. Thus far, these systems have not been able to be reconstructed with the common envelope efficiency of $1/3$ as found by \citet{Zorotovic_2010}, or even with $\alpha_\mathrm{CE} = 1$ although see \citet{Belloni_2024} for a potential solution to this. It is possible for some these to be post-stable mass transfer systems \citep{PathwayIX}, but many of these longer period systems are a challenge to produce even by this channel. Given that WD\,+\,FGK binaries are the last common ancestor to a number of exotic phenomena, such as the before-described cataclysmic variable and double degenerate systems - they are thus important to study and understand. It is possible to find WD\,+\,FGK binaries by looking for sources that appear to be F, G or K type stars in the optical, but have a flux excess in the UV, as a WD would be fully obscured by a F, G or K binary companion in optical wavelengths, but in turn outshine them in the UV owing to their high residual temperatures. Given the importance of such objects, \textit{The White Dwarf Binary Pathways Survey} has set about trying to catalogue the titular systems and determine their formation channels.

\textit{Gaia} data release 3 (hereafter \textit{Gaia}) can be used as an important tool for getting a large number of binary parameters, such as the orbital period, with little effort, as it possesses accurate orbital period measurements for a large number of systems, bypassing the need for extensive follow-up observations to identify and characterise WD\,+\,FGK binaries. In this paper we probe \textit{Gaia} for previously established UV excess binaries (\citealt{PathwayI, PathwayII, PathwayV}) to determine their binary and stellar parameters, most crucially their orbital periods and component masses, so that we can investigate their past and future evolution.

\section{Target Selection}
\label{sec:Targets}

In order to ensure a relatively clean sample of WD\,+\,FGK binaries, we make cuts to remove as many sources of contamination as possible so that we are only working with systems where a WD\,+\,FGK binary is likely, as our method of identifying these systems by their UV excess is only valid for these systems, where the luminous companion completely dominates in optical wavelengths but where the WD causes a notable excess in the UV.

Taking the \textit{RAVE} sample of 430 candidates from \citet{PathwayI}, 1549 candidates from the \textit{LAMOST} sample of \citet{PathwayII} and the sample of 775 candidates from the \textit{TGAS} sample of \citet{PathwayV}, we matched these 2754 candidates to the \textit{Gaia} source catalogue using positional crossmatching. These samples were constructed using \textit{RAVE} data release 4, \textit{LAMOST} data release 4 and \textit{TGAS}; crossmatching them with \textit{GALEX} UV data in order to identify candidate UV excess sources.

To this point, these samples may contain poor \textit{Gaia} matches and stars that are not of the F, G or K spectral classes. To resolve this, we implemented a cut to our selection criteria - removing systems that were a magnitude bluer than the main sequence track as defined by the {\sc MIST}\footnote{\url{https://waps.cfa.harvard.edu/MIST/}} isochrone \citep{MIST0,MIST1,MESA_Stel_Astro,MESA_PORMS,MESA_BPE} for a solar metalicity star using \textit{Gaia} \textit{G} and \textit{G$_\mathrm{BP}$} - \textit{G$_\mathrm{RP}$} magnitudes, which removed 6 systems from the \textit{RAVE} sample of \citet{PathwayI}, 204 from the \textit{LAMOST} sample of \citet{PathwayII} and 1 systems from the \textit{TGAS} sample of \citet{PathwayV}. For the astrometric binaries, we used the parallax values from the \textsc{gaiadr3.nss\_two\_body\_orbit} catalogue throughout this paper. This removed sources where the compact object may contribute to the optical flux, such as WD\,+\,dM or hot subdwarf\,+\,FGK binaries. Whilst measures were taken by \citet{PathwayI}, \citet{PathwayII} and \citet{PathwayV} to remove M stars, some have since been flagged by \textit{Gaia}, while no efforts were taken to remove hot subdwarf systems in the original studies.

\begin{figure*}
    \centering
    \includegraphics[height=0.9\textheight]{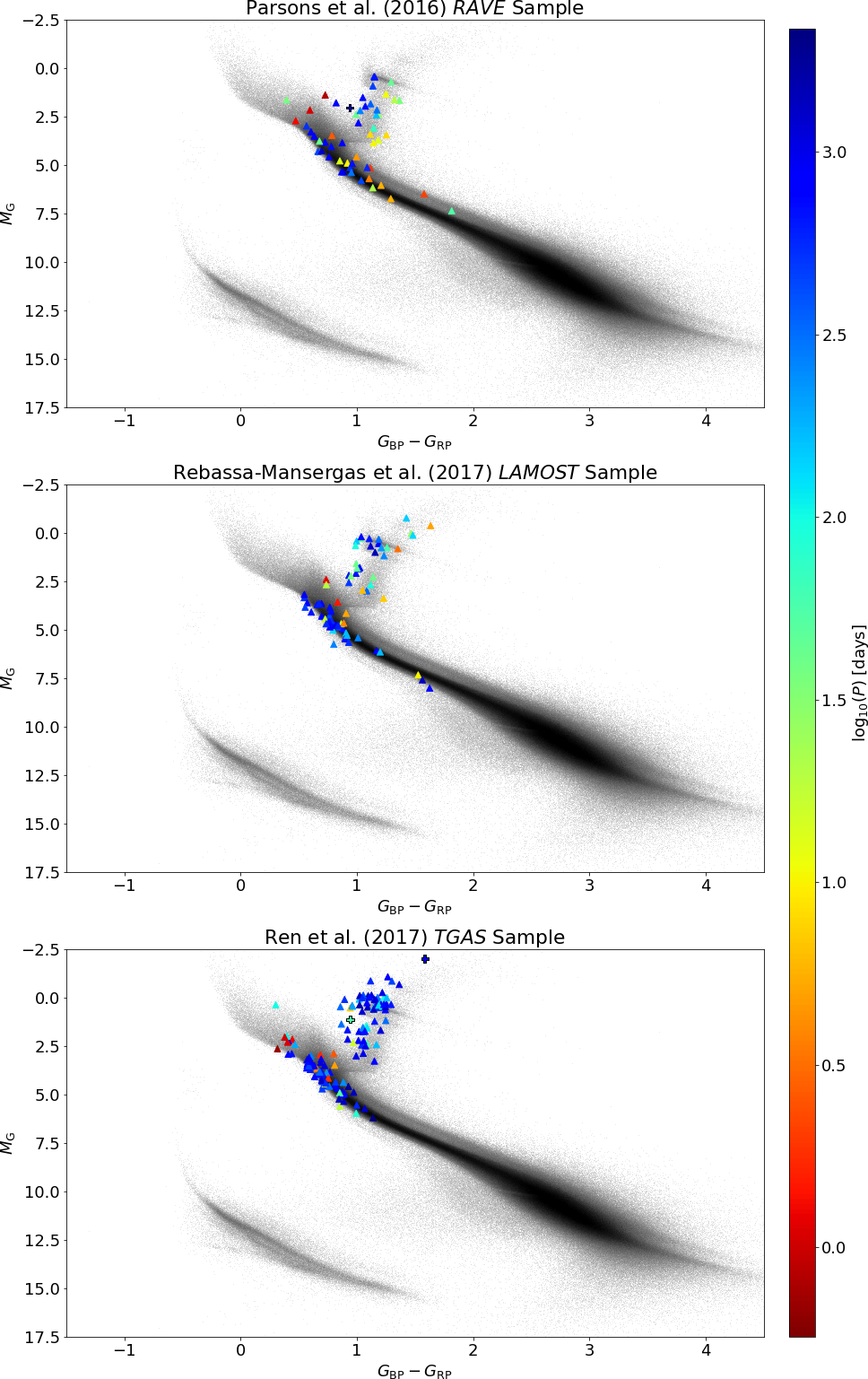}
    \caption{Optical colour-magnitude diagrams using \textit{Gaia} $G$, $BP$ and $RP$ magnitudes, split by the three surveys of \citet{PathwayI}, \citet{PathwayII} and \citet{PathwayV}. The points are colour-coded by the log\textsubscript{10} of their period. The background points are a randomly selected sample of stars from \textit{Gaia} within 250~pc and with a parallax-over-error greater than 20. The larger, circular points represent candidate hot subdwarf binary systems (see Section~\ref{sec:sdOB}).}
    \label{fig:Optical_HRs}
\end{figure*}

Following the removal of these contaminants, along with the removal of duplicates within each survey, we crossmatched the surveys with the {\sc gaiadr3.nss\_two\_body\_orbit} catalogue, which provided us with orbital solutions for systems identified as astrometric, spectroscopic or eclipsing binaries. The optical colour-magnitude diagrams of each of the surveys, colour-coded by their orbital period, can be observed in Figure~\ref{fig:Optical_HRs}. It is worth noting that the presence of giant stars (the outcrop of points on the luminous, redder end of the main sequence track) in the \textit{RAVE} and \textit{LAMOST} samples was somewhat unexpected, as each had taken measures to avoid giant stars, with the other points sitting between $\sim\,0.3$ and $\sim\,1.8$ in the $G_\mathrm{BP}\,-\,G_\mathrm{RP}$ domain, roughly where we would expect for F, G and K stars along the main sequence track. We can see that, in contrast to previous results, there are a large number of long-period systems\footnote{We refer to systems with orbital periods in the order of $\sim$\,100-1000 days as long-period binaries, as opposed to wide binaries which have periods of many years and will not have had any binary interactions.} found across all three samples, with \textit{TGAS} especially being dominated by periods in excess of 100 days. It can also be observed that the \textit{TGAS} sample does not stretch as far down the main sequence track as much as either the \textit{RAVE} or \textit{LAMOST} samples, which could be down to selection bias, as \citet{PathwayV} was conducting a sample of WD\,+\,AFGK binaries opposed to just WD\,+\,FGK binaries, which may have skewed their selection away from the fringe cases of late-K stars, or owing to the initial selection that used DR1 as the original sample.

The solutions found within the {\sc gaiadr3.nss\_two\_body\_orbit} catalogue are: `Orbital', `OrbitalAlternative', `OrbitalTargetedSearch',  `OrbitalTargetedSearchValidated' (all four of which are astrometric fits, and will hereafter be referred to as astrometric systems), `EclipsingBinary', `EclipsingSpectro' (which are both visually eclipsing systems), `SB1', `SB2', `SB1C', `SB2C' ( which are all spectroscopic systems - the number referring to how many resolved lines are present, and the C indicating a circular orbit), and `AstroSpectroSB1' (which is a combined astrometric and single-line spectroscopic model, but for our purposes can be counted as solely astrometric). The breakdown of how many systems belonged to each solution type is detailed in Table~\ref{tab:Sol_Type}, along with the number of eclipsing binaries found by cross-matching with the \textsc{gaiadr3.vari\_classifier\_result} catalogue.

\begin{table}
    \centering
    \caption{A table breaking down the different solution types found across \citet{PathwayI}, \citet{PathwayII} and \citet{PathwayV} (given the short-hand form of of \textit{RAVE}, \textit{LAMOST} and \textit{TGAS} respectively). The numbers given outside the brackets are the numbers in our final sample of WD\,+\,FGK candidates, whilst those given in the brackets represent the number found within the \textsc{gaiadr3.nss\_two\_body\_orbit} catalogue (or flagged as ECL in the \textsc{gaiadr3.vari\_classifier\_result catalogue}. The total number of candidates here listed is 211, with the discrepancy between this number and our quoted number of 206 candidates arising from duplicates between the samples. Additionally, the total number of objects stated here from the {\sc gaiadr3.nss\_two\_body\_orbit} catalogue occur before our listed quality control cuts.}
    \begin{tabular}{llll}
    \hline
        Solution / Systems & \textit{RAVE} & \textit{LAMOST} & \textit{TGAS} \\

        \hline

        Orbital & 4\,(7) & 24\,(30) & 18\,(26) \\

        OrbitalTargetedSearch & 0 & 0 & 1\,(2) \\

        OrbitalTargetedSearchValidated & 0\,(1) & 0 & 0 \\

        SB1 & 25\,(37) & 25\,(33) & 56\,(73) \\

        SB2 & 0 & 0 & 0\,(6) \\

        SB2C & 0 & 0 & 0\,(4) \\

        AstroSpectroSB1 & 12\,(14) & 8\,(9) & 38\,(42) \\

        Eclipsing Binaries & 0\,(0) & 0\,(1) & 0\,(4) \\

        Total & 41\,(59) & 57\,(72) & 113\,(153)
        
    \end{tabular}
    \label{tab:Sol_Type}
\end{table}

With the solution types known, we can make further cuts in accordance with \citet{Gaia_Stellar_Multiplicity}. Which is to say `Orbital' systems were excluded if;

\begin{itemize}
    \item {\sc phot\_g\_mag} $>$ 19,
    \item {\sc ipd\_frac\_multi\_peak} $\geq$ 2,
    \item {\sc ipd\_gof\_harmonic\_amplitude} $>$ 0.1,
    \item {\sc visibility\_periods\_used} $<$ 11,
    \item $C^{*}$ $<$ 1.645$\sigma_\mathrm{C^{*}}$\footnote{Here, $C^{*}$ is the corrected $BP$ and $RP$ flux excess and $\sigma_\mathrm{C^{*}}$ is its associated uncertainty, as defined by \citet{Gaia_Photometric}.},
\end{itemize}

where {\sc ipd\_frac\_multi\_peak} is the percent of successful Image Parameter Determination (IPD) with more than one peak, {\sc ipd\_gof\_harmonic\_amplitude} is the amplitude of the the IPD goodness of fit vs the postition angle of the associated scan, and {\sc visibility\_periods\_used} is the number of visibility periods used in the astrometric solution. For `SB1' and `SB1C', the criteria for exclusion were;

\begin{itemize}
    \item {\sc rv\_renormalised\_gof} < 4,
    \item {\sc rv\_nb\_transits} < 11,
    \item 3875 > {\sc rv\_template\_teff} > 8125,
\end{itemize}

where {\sc rv\_renormalised\_gof} is the renormalised goodness of fit of the radial velocity measurements, {\sc rv\_nb\_transits} is the number of transits used in the calculation of the radial velocity, and {\sc rv\_template\_teff} is the effective temperature ($T_{\mathrm{eff}}$ of the template used in the radial velocity calculations.

Since we are interested in systems where the optical flux comes exclusively from one star, systems flagged as `EclipsingBinary', `EclipsingSpectro', `SB2' or `SB2C' were dropped from the samples, as the Eclipsing systems by their nature contain  stars bright enough to show an eclipse (which would not be possible except for very hot WDs with very late-K stars), and the double-lines spectroscopic by their nature have two optically luminous components, which would not be the case for a WD\,+\,FGK binary. Additionally, systems which were flagged as `ECL' (shorthand for eclipsing) in the {\sc gaiadr3.vari\_classifier\_result} were likely dropped, for the same reason as stated for the eclipsing systems within {\sc gaiadr3.nss\_two\_body\_orbit}. After these systems are removed and the above astrometric and spectroscopic cuts are applied, we are left with 55 systems from the \textit{RAVE} sample of \citet{PathwayI}, 71 from the \textit{LAMOST} sample of \citet{PathwayII} and 128 from the \textit{TGAS} sample of \citet{PathwayV}. It should be noted that there are systems that lie across multiple of these surveys, with there being a total of 246 systems after this criterion is taken into account.

For the astrometric binaries, we used the python {\sc nss tools}\footnote{\url{https://gitlab.obspm.fr/gaia/nsstools}} \citep{Gaia_Processing} in order to transform the given orbital Thiele-Innes constants, denoted as $A$, $B$, $F$ and $G$, and defined by;
\begin{equation}
    A = a_0 \left[ \cos(\omega) \cos(\Omega) - \sin(\omega) \sin(\Omega) \cos(i)\right],
\end{equation}
\begin{equation}
    B = a_0 \left[ \cos(\omega) \sin(\Omega) + \sin(\omega) \cos(\Omega) \cos(i)\right],
\end{equation}
\begin{equation}
    F = -a_0 \left[ \sin(\omega) \cos(\Omega) + \cos(\omega) \sin(\Omega) \cos(i)\right],
\end{equation}
\begin{equation}
    G = -a_0 \left[ \sin(\omega) \sin(\Omega) - \cos(\omega) \sin(\Omega) \cos(i)\right],
\end{equation}
where $a_0$ is the angular semi-major axis of the system, $\omega$ is the argument of periapsis, $\Omega$ is the longitude of the ascending node and $i$ is the inclination; into the Campbell parameters ($a_0$, $\omega$, $\Omega$ and $i$), the most pertinent of which being the angular semi-major axis, $a_0$ of the system. 

\section{Stellar Parameters}
\label{sec:Parameters}

With our samples cleaned and orbital parameters acquired, we can set about obtaining the parameters of the luminous star\footnote{Luminous star here referring to the more optically luminous companion, which can be either a main sequence or an evolved star.} and then the WD. This will allow us to probe the evolution of these systems, along with flagging additional contaminants.

\subsection{Luminous star parameters}
\label{sec:LSParams}

Whilst \textit{Gaia} does give estimates for stellar parameters in their {\sc gaiadr3.astrophsical\_parameters}, {\sc gaiadr3.astrophsical\_parameters\_supp} and {\sc gaiadr3.binary\_masses} catalogues, these are not necessarily accurate and are not available for all systems. As such, we obtain properties for the luminous component of the binary system using spectral energy distribution (SED) fitting with the python {\sc Isochrones} package\footnote{\url{https://github.com/timothydmorton/isochrones}} \citep{Isochrones_Package}. We perform fitting using synthetic $u$, $g$, $r$, $i$, $z$ magnitudes from {\sc gaiadr3.synthetic\_photometry\_gspc} where available, $g$, $r$, $i$, $z$ if there was no $u$ measurement, or $G$, $BP$ and $RP$ as a last resort if there was no synthetic photometry available - alongside $J$, $H$, $Ks$, $W1$ and $W2$ magnitude measurements taken from \textit{2MASS} \citep{2MASS} and \textit{AllWISE} \citep{AllWISE} \textit{Gaia} match catalogues ({\sc gaiadr1.tmass\_original\_valid} with {\sc gaiadr3.tmass\_psc\_xsc\_best\_neighbour} and {\sc gaiadr1.allwise\_original\_valid} with {\sc gaiadr3.allwise\_best\_neighbour} respectively). SED fitting is not without its drawbacks - it relies on assumptions about its inputs and our results without $u$ band synthetic photometry in particularly may not be able to constrain the effective temperature, $T_\mathrm{eff, LS}$ very well. Ultimately, spectroscopic data is needed to fix these parameters properly.

We place priors on our fit, these include a Gausian prior on parallax based on \textit{Gaia} measurements, a flat prior on extinction, based on the values and uncertainties from the 3d dustmaps of \textsc{stilism} \citet{Stilism}, with a Jeffreys prior, as defined by
\begin{equation}
    P(E(B-V) \propto E(B-V)^{1/2}.
\end{equation}
where we place a lower bound of $10^{-4}$ and an upper bound at the maximum value found in the system's direction within stilism. We also used the \citet{Casagrande_2011} metalicity prior. We then fitted our data to \textsc{mist} isochrones \citep{MIST0,MIST1,MESA_Stel_Astro,MESA_PORMS,MESA_BPE} using a Markov Chain Monte Carlo (MCMC) method with 3000 live points per fit. For our initial fits we assumed that all the flux comes from the optically luminous star (hence why we did not include UV measurements in our fit), however there may be cases where the WD component could contribute to the optical flux, particularly at the shortest wavelengths. Using the WD parameters derived in Section~\ref{sec:WDParams}, we can estimate what the WD contribution is likely to be at optical wavelengths and found that this reaches around 5 per cent at most. We ran additional fits to re-derive the luminous star parameters accounting for the WD contribution and found that even in the worst cases the resulting difference in the parameters was far smaller than the uncertainties on the original measurements.

Some non-WD companions (usually low mass active stars) can contaminate UV excess samples such as the ones we are dealing with. These contaminants can be found by looking for infrared excesses, as a low-mass star will contribute a non-negligible amount of flux at longer wavelengths. To identify these kinds of contaminants, we fitted an SED with a two component fit using the {\sc Isochrones} package. We then performed a log-likelihood ratio test on every system comparing their single and double star model fits, flagging systems where the double fit was more likely (comprising 23 systems) as likely contaminants. We also note that it is possible to hide relatively high mass main-sequence companions (e.g. A-type stars) next to giants, which may result in the detection of a UV excess, while having little effect on the optical colours of the system. However, these stars have masses typically much higher than the Chandrasekhar mass and so can be removed with a simple upper limit on the companion mass. Ultimately, of the combined sample of 246 systems, we have 43 contaminants, giving us a contamination rate of $\sim$\,17 per cent, which is fairly consistent with the previous works in this series.

This gave us values describing the luminous companion - its effective temperature, the log of its surface gravity, log($g$)$_\mathrm{LS}$, its radius, $R_\mathrm{LS}$, and most importantly, its mass, $M_\mathrm{LS}$, all of which can be seen in appendix~\ref{sec:tables_of_Params}. It also provided values for metallicity, log($Z$)$_\mathrm{LS}$, though the SED is insensitive to this parameter and is in many cases unconstrained, so we do not include it here.

\subsection{White dwarf parameters}
\label{sec:WDParams}

By using the parameters of the luminous companion determined by SED fitting with the \textit{Gaia} orbital parameters, we can determine the properties of the hidden companion, assumed to be a WD. This is done using a different approach depending on wether the system is an astrometric binary or a spectroscopic binary, so they will be discussed separately.

\subsubsection{Astrometric binaries}

We can determine the mass of the unseen companion within an astrometric binary using the astrometric mass function, $\mathcal{A}$, described by \citet{Shahaf_2019}, 
\begin{equation}
\label{equ:AstroMass}
    \mathcal{A} \equiv \frac{q}{(1 + q)^{\frac{2}{3}}} \left( 1 - \frac{\mathcal{S}(1 + q)}{q(1 + \mathcal{S})} \right) = \frac{a_0}{\varpi} {M_{\mathrm{LS}}^{-\frac{1}{3}} P^{-\frac{2}{3}}},
\end{equation}
in which $a_0$ is the angular semi-major axis, $\varpi$ is the parallax, $P$ is the period, $q$ is the mass ratio of the system, such that
\begin{equation}
    q = \frac{M_\mathrm{WD}}{M_\mathrm{LS}},
\end{equation}
and $\mathcal{S}$ is the ratio of the intensities of the components, which we take as zero as our underlying assumption is that 100 per cent of the optical flux is contributed by the luminous companion.

It is thus straightforward to extract the mass of the WD by solving the astrometric mass function. We also propagated all uncertainties through this function to estimate errors on the WD masses. Any systems where the unseen companion has a mass clearly above the Chandrasekhar mass limit were flagged as contaminants.

\subsubsection{Spectroscopic binaries}

Spectroscopic binary systems are instead determined by the binary mass function determined from Kepler's laws, 
\begin{equation}
    \frac{{M_{\mathrm{WD}}^{3} \sin^{3}(i)}}{(M_{\mathrm{WD}} + M_{\mathrm{LS}})^{2}} = \frac{K^{3} P}{2 \pi G} (1 - e^{2})^{\frac{3}{2}},
\end{equation}
where $K$ is the radial velocity semi-amplitude and $e$ is the eccentricity. It should be noted that in general the inclination, $i$, is unknown for spectroscopic systems. This limits us to only getting a minimum WD mass estimate for these systems by assuming $\sin{i}=1$, which will limit the conclusions that can be drawn from these systems.

\subsubsection{White dwarf temperature estimates}
\label{sec:WDTeff}

With the masses known (or lower limits in the case of the spectroscopic systems), and $FUV$ magnitudes measured by \textit{GALEX} \citep{GALEX}, we can set about obtaining estimates of the effective temperature of the WD $T_\mathrm{eff, WD}$ (and consequently the log of its surface gravity, log($g$)$_\mathrm{WD}$) and its cooling age, $\tau_\mathrm{cool}$. This allows us to estimate how long it has been since the last mass transfer phase, which also tells us if the luminous star is likely to be out of thermal equilibrium (i.e. if its thermal timescale is longer than the WD cooling age). The WD cooling age is particularly important for properly understanding the evolution of the shortest period systems, where additional angular momentum loss may have significantly altered their orbital period from what it was after the last mass transfer phase. It is important to correct for this effect when reconstructing their evolution \citep[e.g.][]{Zorotovic_2010}.

In order to obtain these values, we matched our de-reddened $FUV$ values (obtained by taking the $FUV$ magnitude from \textit{GALEX} and de-reddening by our calculated reddening values) and our WD mass values against interpolated values from models. For WDs with a mass measured below 0.5~M\textsubscript{$\odot$} we used the helium core models of \citet{Althaus_2013}\footnote{\url{http://evolgroup.fcaglp.unlp.edu.ar/TRACKS/newtables.html}}, whilst for more massive WDs (above 0.5~M\textsubscript{$\odot$}) we used the carbon/oxygen core models of \citet{Bédard_2020}\footnote{\url{https://www.astro.umontreal.ca/~bergeron/CoolingModels/}}, assuming a standard mass-radius relationship. Note that, as our spectroscopic masses are only lower limit estimates, this likewise puts a lower limit on our $T_\mathrm{eff, WD}$ values (since if the WD has a higher mass, hence a smaller radius, its temperature must increase to match the UV flux). Since some of our spectroscopic systems have minimum WD masses below 0.2~M\textsubscript{$\odot$} we are not able to get temperature estimates for all our systems.

With our stellar and binary parameters established (which can be seen in appendix~\ref{sec:tables_of_Params}), we can compare our values to those found by other means to check the integrity of our method, and investigate the implications of what our parameters tell us about the evolution of these systems. 

Based on our $T_\mathrm{eff, WD}$ values, we can estimate the WD flux contributions in various passbands. We find that in the $u$ band for example the vast majority of WDs contribute less than 5 per cent of the observed flux, and in most cases less than 1 per cent. This is true in the $G$ band, which \textit{Gaia} uses for astrometric measurements, and is therefore relevant to equation~\ref{equ:AstroMass}, in particular $\mathcal{S}$, the ratio of intensities. However, even accounting for a worst case scenario of a 5 per cent contribution in the $G$ band ($\mathcal{S} = 0.05$), the resulting increase in $M_{\mathrm{WD}}$ is smaller than the uncertainties from our method.

\section{Hubble Space Telescope Observations}

Eight of our systems have been observed spectroscopically at UV wavelengths with the \textit{Hubble Space Telescope (\textit{\textit{HST}})}. These data allow us to confirm the origin of the UV excess in these objects and in the cases where this is due to a WD, we can determine their physical parameters by fitting the spectrum. Here we present \textit{HST} data for four new objects comprising three WD\,+\,FGK binaries and an active star (TYC~6086-1317-1, TYC~9151-303-1, TYC~6434-457-1 and CPD-82~849), and refer the reader to previous publications for the other four objects (see section~\ref{sec:CompHubble}), though a breakdown of all eight systems is given briefly in Table~\ref{tab:Hubble_Objects}.

\begin{table*}
    \centering
    \caption{The eight systems within our sample which have \textit{HST} UV spectra, with the Cosmic Origins Spectrograph (COS) using the G130M grating centred on 1291\,{\AA} and the Space Telescope Imaging Spectrograph (STIS) using the MAMA detector and the G140L grating centred on 1425\,{\AA}. Listed are the proposed source of their UV excess, the compiled exposure time and the source of publication.}
    \tabcolsep=0.15cm
    \begin{tabular}{@{}lccccc@{}}
        \hline 
        
        Name & UV Source & Observation Date & Exposure time [s] & Source \\

        \hline

        TYC~9151-303-1 & WD in binary & 2015\,/\,07\,/\,14 & 2441 & This paper\\
        
        CPD-82~849 & Active star & 2021\,/\,04\,/\,05 & 8718 & This paper\\

        TYC~6086-1317-1 & WD in binary & 2021\,/\,03\,/\,01 & 2196 & This paper\\

        TYC~6434-457-1 & WD in binary & 2020\,/\,12\,/\,31 & 4934 & This paper\\

        2MASS~J06281844-7621467 & Active star & 2021\,/\,04\,/\,21 & 11731 & \citet{PathwayVII} \\
        
        TYC~7218-934-1 & WD companion to MS\,+\,MS binary & 2015\,/\,04\,/\,29& 4384 & \citet{PathwayIII} \\
        
        TYC~6996-449-1 & WD companion to MS\,+\,MS binary & 2014\,/\,12\,/\,03 & 2418 & \citet{PathwayVII} \\

        TYC~6992-827-1 & WD in binary & 2022\,/\,01\,/\,10 & 4980 & \citet{PathwayIX} \\

        \hline
    \end{tabular}
    \label{tab:Hubble_Objects}
\end{table*}

For the four \textit{HST} spectra clearly containing a WD we fitted synthetic DA WD models in order to determine the stellar parameters. We followed the procedure outlined in \citet{PathwayIX}, whereby we used a pre-generated model grid created using the code of \citet{Koester2010}, spanning a range of effective temperatures of 12000--30000\,K in steps of 200\,K and surface gravities of 6.0-9.0 in steps of 0.1 dex, and interpolate between these grid points. Additionally, we included the effects of reddening and interstellar neutral hydrogen (although the latter has a very minor effect) and scale the model flux based on parallax. Thus the free parameters were effective temperature, surface gravity, parallax and reddening. WD masses, radii and cooling ages are then determined via a theoretical mass-radius relationship (\citealt{Bedard2020} for masses above 0.5\,M$_{\odot}$ or \citealt{Althaus_2013} for masses below 0.5\,M$_{\odot}$).

We used the Markov Chain Monte Carlo (MCMC) method to fit the spectra, implemented using the ensemble sampler {\sc emcee} \citep{Foreman2013}, with 4000 steps and 100 walkers, where the autocorrelation time for each parameter was generally found to be less than 100 steps, and a burn-in period of 5 times that of the autocorrelation time was used. We applied Gaussian priors to the parallax and reddening based on measurements from \textit{Gaia} and the 3D reddening map of {\sc stilism} \citep{Stilism}, respectively. We also add in quadrature systematic errors of 1.5 per cent in $T_\mathrm{eff}$ and 0.04~dex in log($g$) \citep{PathwayIX}. For the astrometric binary systems we also generated model spectra based on the dynamically determined WD parameters.

\section{Discussion}
\label{sec:Discussion}

\subsection{Comparison of dynamical white dwarf masses to spectroscopic masses from \textit{HST} UV spectroscopy}
\label{sec:CompHubble}

\begin{figure}
    \centering
    \includegraphics[width=\columnwidth]{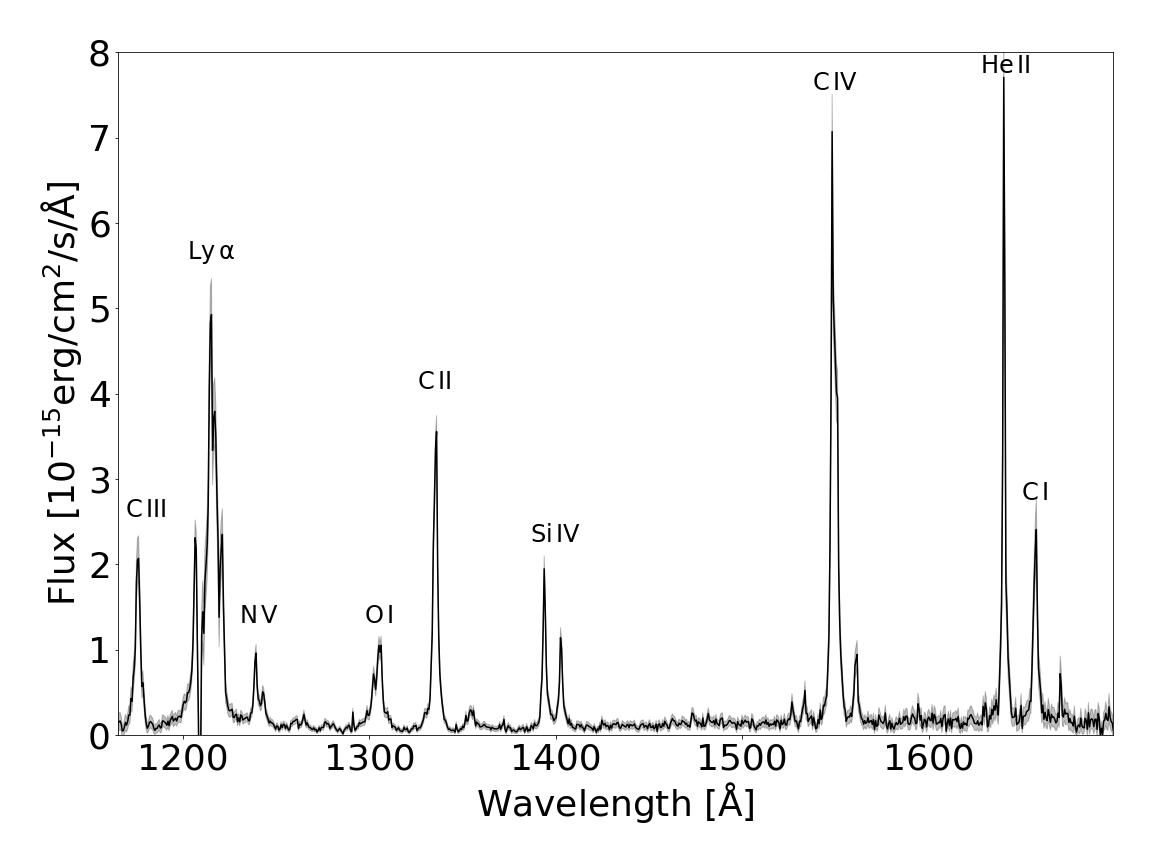}
    \caption{The \textit{HST}/STIS spectrum of CPD-82~849. The spectrum lacks any clear indication of a WD, but does show strong emission lines typically seen in active stars, which are likely the origin of the UV excess in this system.}
    \label{fig:HST_CPD_82_849}
\end{figure}

Of the eight systems previously mentioned to have been observed with \textit{HST} spectroscopy, four (2MASS~J06281844-7621467, TYC~7218-934-1, TYC~6996-449-1 and CPD-82~849) were identified as contaminants, with CPD-82~849 having not yet been published. Its UV spectrum (as seen in Figure~\ref{fig:HST_CPD_82_849}) clearly identifies it as an active star.

\begin{table*}
    \centering
    \caption{A comparison of the WD parameters found through orbital analysis and SED fitting of \textit{Gaia} data ("dynamical") against those from fitting \textit{HST} UV spectra ("spectroscopic"). The analysis of the UV spectrum of TYC~6992-827-1 was presented in \citet{PathwayIX} and we list their values here. Both TYC~6992-827-1 and TYC~6434-457-1 are spectroscopic binaries, hence the dynamical constraints represent lower limits (i.e. assuming $\sin(i)=1$), which for TYC~6992-827-1 gives a WD mass so low that we are unable to estimate its temperature or surface gravity. The two methods give consistent results to within 2-3 $\sigma$ in all cases.}
    \tabcolsep=0.15cm
    \begin{tabular}{@{}lllccccc@{}}
        \hline 
        
        Name & P$_\mathrm{orb}$ [d] & Method & $T_\mathrm{eff, WD}$ [K] & log($g$)$_\mathrm{WD}$ [dex] & $\tau_\mathrm{cool}$ [Myr] & $M_\mathrm{WD}$ [M\textsubscript{$\odot$}]\\

        \hline

        TYC~6992-827-1 & 41.45 $\pm$ 0.01 & Dynamical & - & - & - & $>$0.13 $\pm$ 0.01 \\
        
        & & \citet{PathwayIX} & $15750 \pm 50$ & 7.14 $\pm$ 0.02 & 6.9 $\pm$ 0.1 & 0.28 $\pm$ 0.01 \\

        TYC~6086-1317-1 & 524.2 $\pm$ 3.3 & Dynamical & 20100 $\pm$ 400 & 7.29 $\pm$ 0.04 & 40 $\pm$ 6 & 0.38 $\pm$ 0.01 \\
        
         & & Spectroscopic & 19850 $\pm$ 350 & 7.12 $\pm$ 0.05 & 20 $\pm$ 12 & 0.34 $\pm$ 0.02 \\

        TYC~9151-303-1 & 797.8 $\pm$ 9.6 & Dynamical & 15700 $\pm$ 300 & 7.51 $\pm$ 0.03 & 190 $\pm$ 13 & 0.42 $\pm$ 0.01\\
        
         & & Spectroscopic & 14900 $\pm$ 250 & 7.35 $\pm$ 0.06 & 184 $\pm$ 26 & 0.36 $\pm$ 0.03 \\
        
        TYC~6434-457-1 & 930.2 $\pm$ 9.8 & Dynamical & $>$14800 $\pm$ 1500 & $>$7.78 $\pm$ 0.12 & $>$234 $\pm$ 26 & $>$0.51 $\pm$ 0.06\\
        
         & & Spectroscopic & 13600 $\pm$ 300 & 7.93 $\pm$ 0.12 & 252 $\pm$ 23 & 0.58 $\pm$ 0.03 \\

        \hline
    \end{tabular}
    \label{tab:Hubble_Comp}
\end{table*}

The other four systems with \textit{HST} spectra - TYC~6992-827-1, TYC~6086-1317-1, TYC~9151-303-1 and TYC~6434-457-1 - clearly contain a WD, with the best-fit parameters shown in Table~\ref{tab:Hubble_Comp}, and are each discussed in more detail below. For the following discussion we refer to the parameters measured via the \textit{Gaia} orbital solution plus {\it GALEX/FUV} fit as "dynamical" measurements, while those measured via fitting the {\it HST} UV spectrum are referred to as "spectroscopic" measurements. It is also worth noting that TYC~9151-303-1 was flagged by our SED fitting method as a contaminant, with the two-star model SED fitting the system better -  which could indicate either that our SED contamination detection method may not always be reliable or that the SED is genuinely contaminated, either by a background source or a distant companion to the binary.

\subsubsection{TYC~6992-827-1}

\begin{figure}
    \centering
    \includegraphics[width=\columnwidth]{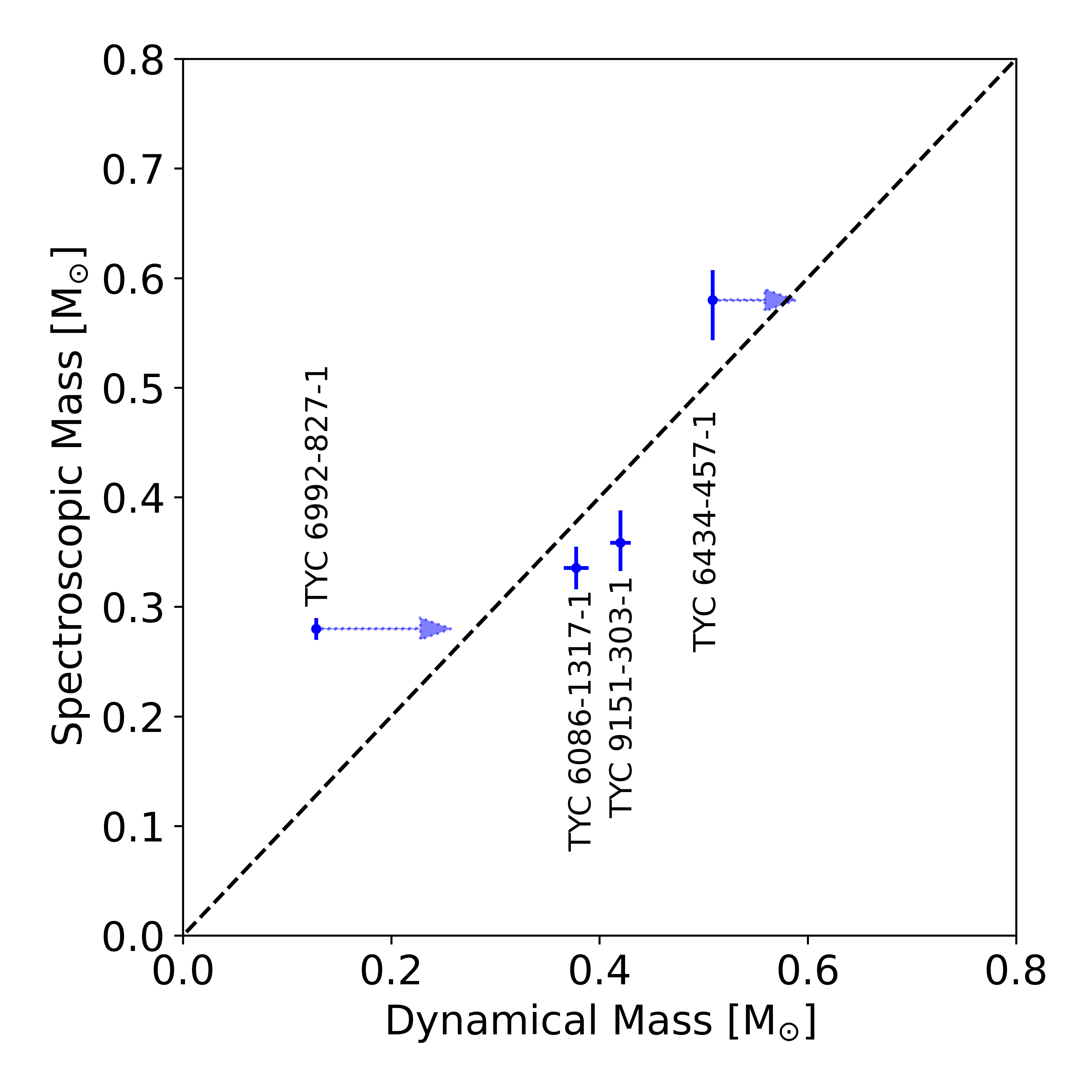}
    \caption{A comparison of the dynamical white dwarf mass from \textit{Gaia} orbits (x-axis) vs those found through \textit{HST} spectral fitting (y-axis).}
    \label{fig:HubbleMass}
\end{figure}

Despite the large difference in mass between our dynamical mass and that of \citet{PathwayIX} shown in Table~\ref{tab:Hubble_Comp}, it should be noted that this system is a spectroscopic binary, and thus this is only a minimum mass estimate. If we are to apply the inclination of $26 \pm 2$\,\textdegree from \citet{PathwayIX}, we instead get a mass of $0.30 \pm 0.02$\,M$_\mathrm{\odot}$, which is consistent with their value, and is illustrated in Figure~\ref{fig:HubbleMass}. This is unsurprising given that the \textit{Gaia} orbital fit is consistent with the ground-based one presented in \citet{PathwayIX}.

\subsubsection{TYC~6086-1317-1}

\begin{figure}
    \centering
    \includegraphics[width=\columnwidth]{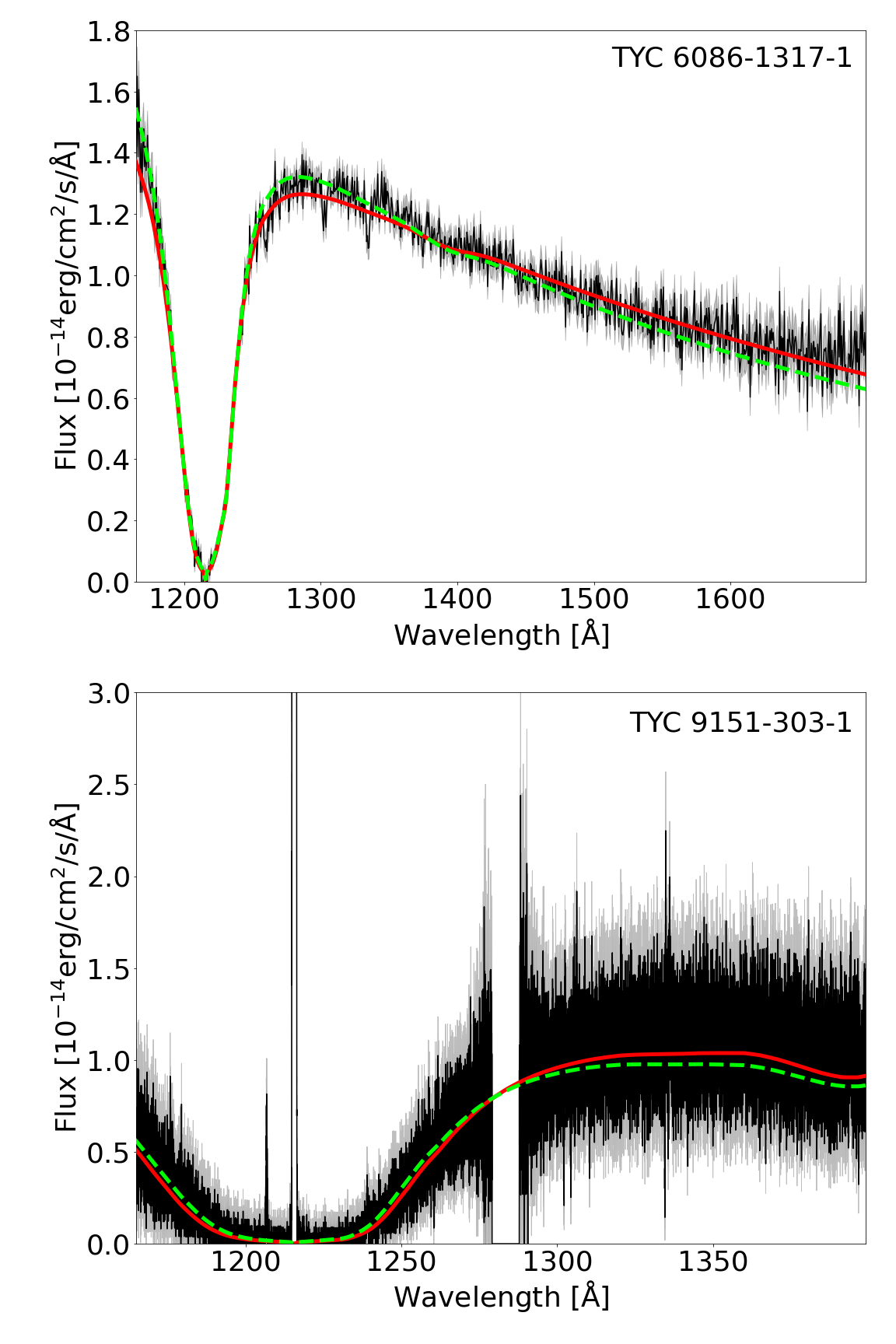}
    \caption{UV spectra of the astrometric binary systems TYC~6086-1317-1 (top, STIS) and TYC~9151-303-1 (bottom, COS). Overplotted are the best fit models in red, while models with the parameters fixed at the "dynamical" values are shown as green dashed lines.}
    \label{fig:HST_Astro}
\end{figure}

TYC~6086-1317-1 was flagged as an astrometric binary system by \textit{Gaia}, and so its mass estimates are more accurate, since the inclination is known. The dynamical and spectroscopic WD parameters are consistent to within 2-3 $\sigma$ demonstrating that the dynamical method can give reliable results. To further highlight this, The \textit{HST} spectrum of this system is shown in the top panel of Figure~\ref{fig:HST_Astro} with both the best fit model spectrum and a model spectrum computed at the dynamically determined WD parameters. Both methods indicate the WD has a low mass and very short cooling age (see Table~\ref{tab:Hubble_Comp}.

\subsubsection{TYC~9151-303-1}

TYC~9151-303-1 was flagged as an astrometric binary system by \textit{Gaia}. Just as with TYC~6086-1317-1 the dynamical and spectroscopic parameters are consistent to within 2-3 $\sigma$. This system also contains a low mass WD, albeit with a slightly longer cooling age than TYC~6086-1317-1. The UV spectrum of this source is shown in the bottom panel of Figure~\ref{fig:HST_Astro}.

\subsubsection{TYC~6434-457-1}

\begin{figure}
    \centering
    \includegraphics[width=\columnwidth]{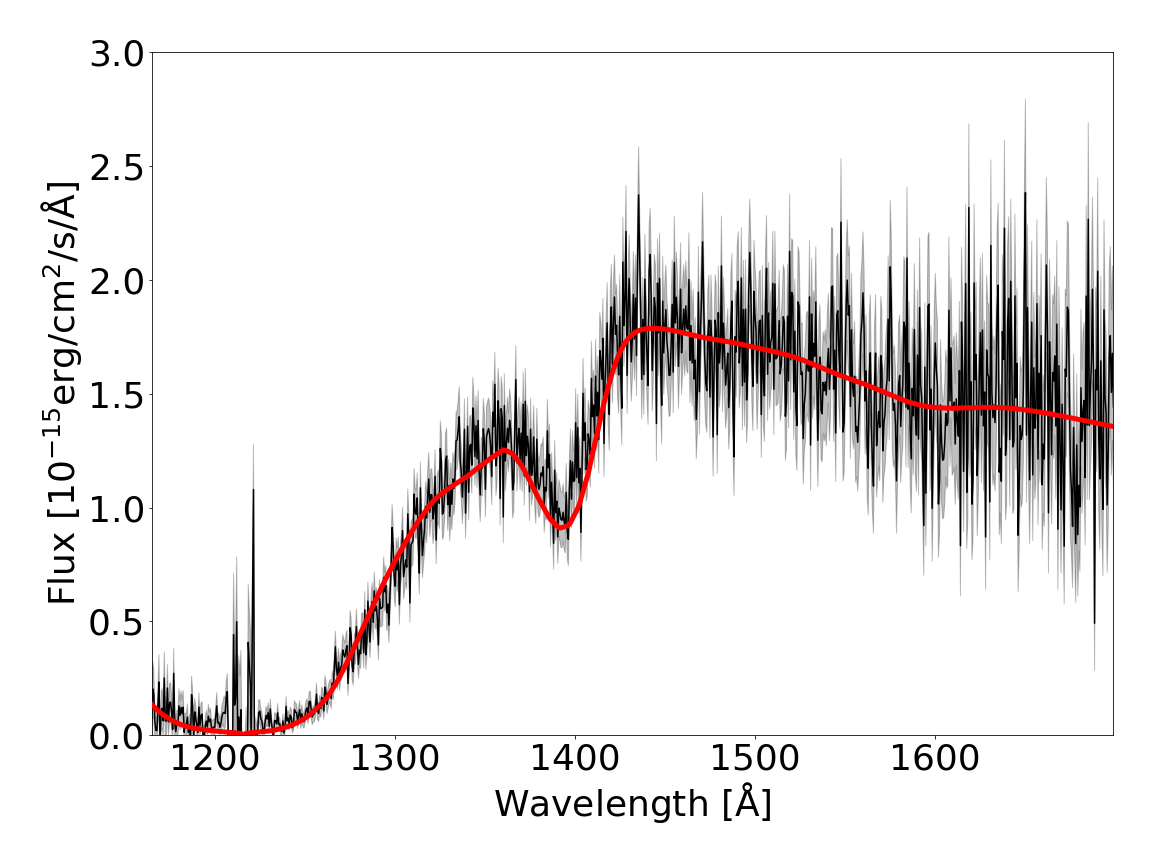}
    \caption{The \textit{HST}/STIS spectrum of the spectroscopic binary system TYC~6434-457-1. The best fit model spectrum in shown in red, we do not plot a model based on the "dynamical" measurements, since this assumes an edge-on orbit and hence under predicts the WD mass.}
    \label{fig:HST_TYC_6434-457-1}
\end{figure}

TYC~6434-457-1 is a single-lined spectroscopic binary in \textit{Gaia}, which has an orbital period comparable to many of the astrometric binary systems. Assuming an edge-on orbit gives a minimum mass for the WD of $0.50 \pm 0.06$ M\textsubscript{$\odot$}, which is somewhat lower than the spectroscopically determined mass of $0.58 \pm 0.03$M\textsubscript{$\odot$}, implying that the system has an inclination of around 60 degrees. In contrast to the two previous objects, TYC~6434-457-1 appears to host a fairly typical mass WD.  The \textit{HST} spectrum of this object is shown in Figure~\ref{fig:HST_TYC_6434-457-1}.

\subsection{Comparing to previous works}
\label{sec:CompRLT}

\begin{figure*}
    \centering
    \includegraphics[width=\textwidth]{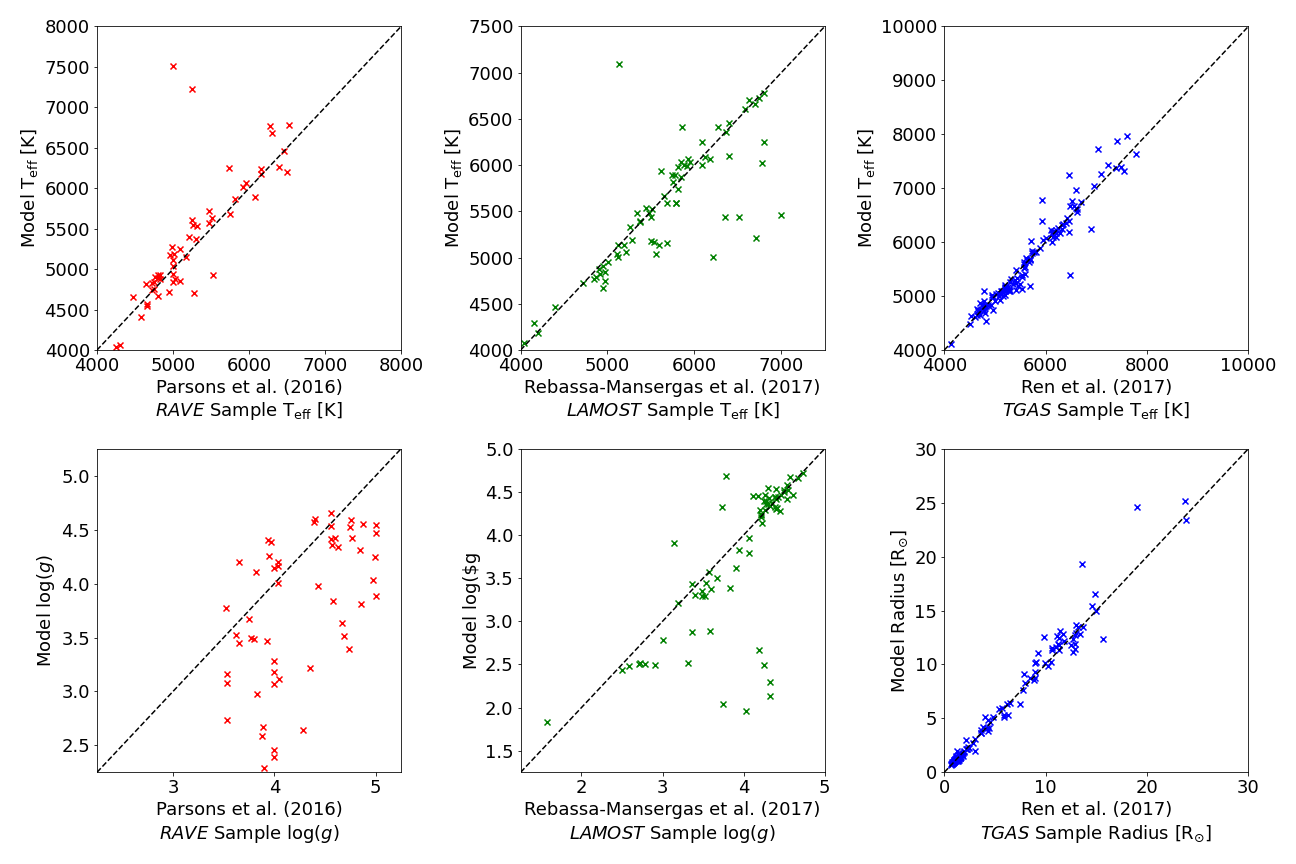}
    \caption{A comparison of our parameters against those of \citet{PathwayI}, \citet{PathwayII} and \citet{PathwayV}. Note that for \citet{PathwayI} and \citet{PathwayII} these were made before any \textit{Gaia} data release, hence the miss-identification of many evolved stars in those studies.}
    \label{fig:Papers_Comp_Plot}
\end{figure*}

Each of the three surveys of \citet{PathwayI}, \citet{PathwayII} and \citet{PathwayV} calculated specific stellar parameters for the luminous companion, with which we can compare our values against. Given that the dynamical WD masses are entirely dependent on the masses of the luminous stars, it is vital that these values are as reliable as possible. \citet{PathwayI} and \citet{PathwayII} each determined values of $T_\mathrm{eff, LS}$ and log($g$)$_\mathrm{LS}$, whilst \citet{PathwayV} measured values for $T_\mathrm{eff, LS}$ and $R_\mathrm{LS}$, the comparison of these values to those found by our dynamical modelling can be observed in Figure~\ref{fig:Papers_Comp_Plot}.

\subsubsection{\textit{RAVE} and \textit{LAMOST}}

We find that, compared to the values of \citet{PathwayI} and \citet{PathwayII}, our temperatures were mostly consistent.

The comparison of our values of log($g$)$_\mathrm{LS}$ is quite telling. There are many systems across both surveys (particularly those of \textit{RAVE}) in which our values were considerably lower than those found by either \citet{PathwayI} or \citet{PathwayII}, which is consistent with our earlier discovery of giant stars in their samples which they had identified as main sequence stars.

\subsubsection{\textit{TGAS}}

Both the temperatures and stellar radii from the survey of \citet{PathwayV} match very closely with the values we found. This is unsurprising, as \citet{PathwayV} likewise obtained their values through SED fitting, though they used parallaxes from \textit{Gaia} data release 2, whilst we used the updated versions from \textit{Gaia} data release 3.

\subsection{Comparing luminous star masses with \textit{Gaia}}
\label{sec:GaiaMass}

\begin{figure}
    \centering
    \includegraphics[width=\columnwidth]{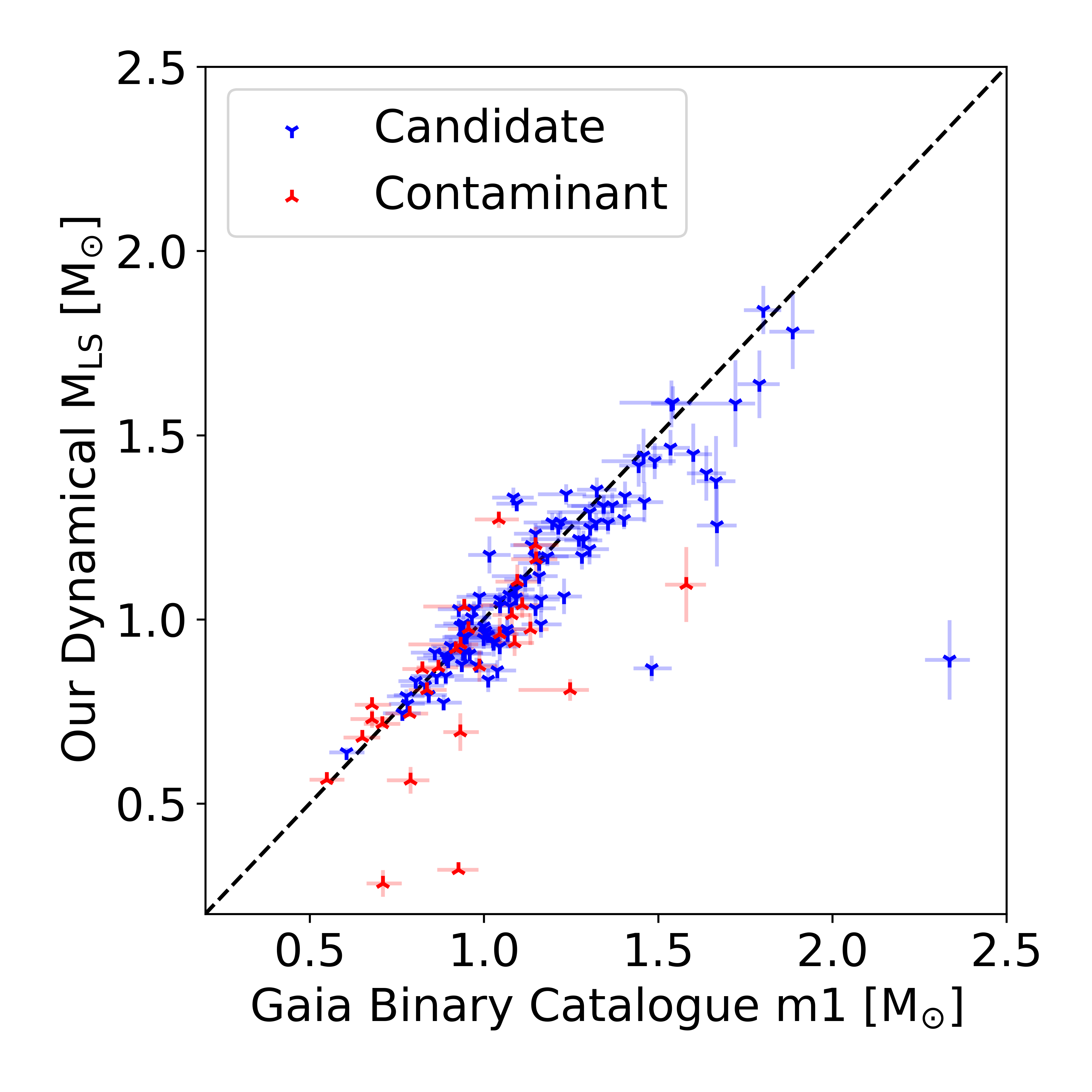}
    \caption{A comparison between our dynamical masses of the more luminous companion against their respective masses found in the {\sc gaiadr3.binary\_masses} catalogue. Only around 50 per cent of our sample had masses in the \textit{Gaia} catalogue, with evolved stars missing from \textit{Gaia}.}
    \label{fig:GaiaMassComp}
\end{figure}

Another useful source to compare against is the {\sc gaiadr3.binary\_masses} catalogue, which contains mass estimates for many systems which \textit{Gaia} has flagged as being binaries. We should note that, of our sample of 246 systems, only 124 had a mass estimates in the \textit{Gaia} binary masses table, as \textit{Gaia} seemingly does not generate parameters for such systems where the optically luminous companion has evolved off the main sequence. Given that this is a crucial parameter to determine the WD mass, there is a clear need to derive these ourselves or WD masses would not be constrained for a significant fraction of our sample. In general, our masses show good agreement with the \textit{Gaia} values, as can be observed in Figure~\ref{fig:GaiaMassComp}, with the more extreme outliers being systems which we have flagged as contaminants.

It should be noted that, of the systems absent from the {\sc gaiadr3.binary\_masses} catalogue, all except 2 can be found as more evolved stars along the giant branch. When using the \textit{Gaia} binary masses table one should be aware of this bias.

\subsection{Evolutionary states}
\label{sec:EvSt}

\begin{figure*}
    \centering
    \includegraphics[width=\textwidth]{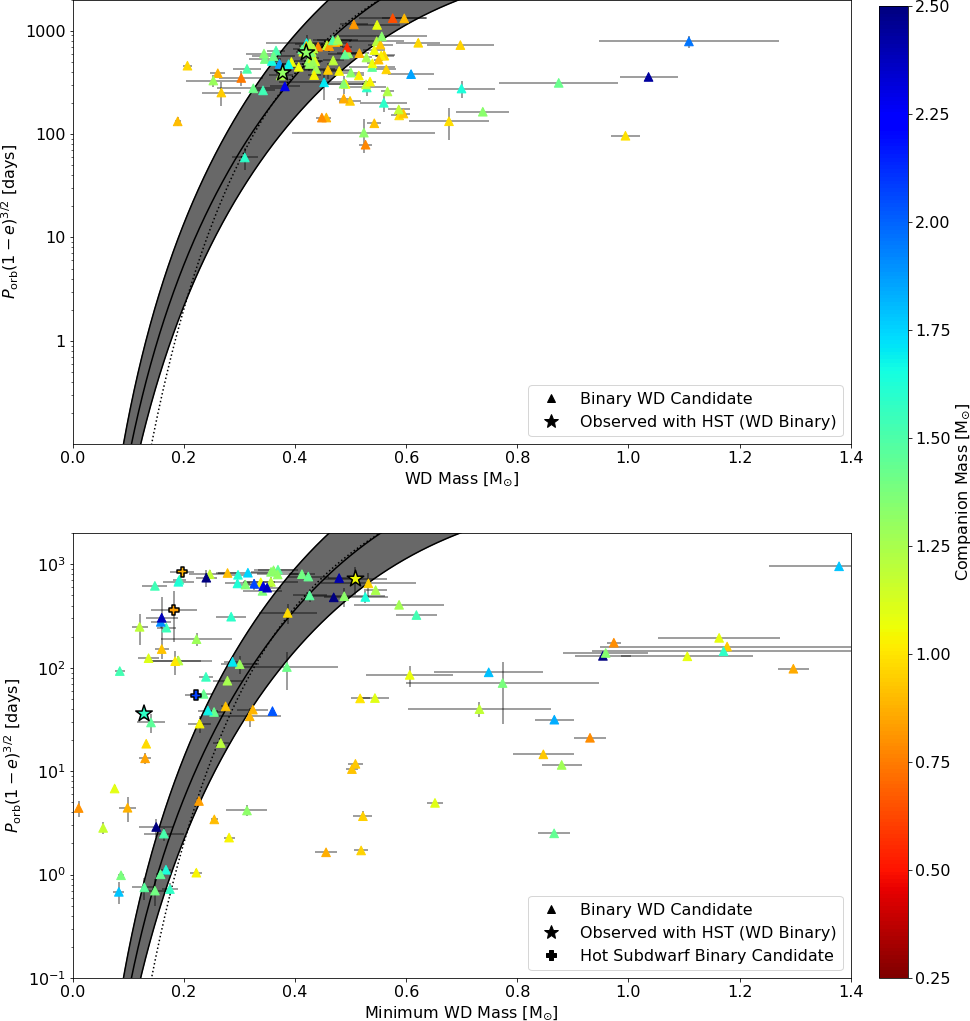}
    \caption{$M_\mathrm{WD}$/$P$ relationships for our candidate WD binary systems, colour-coded by the mass of the luminous companion, and split by their solution type (top panel: astrometrically detected systems, bottom panel: spectroscopically detected systems). Of note, the WD mass is only a minimum mass estimate for the spectroscopic systems. The solid black line and shaded grey region represent the theoretical region for conservative post-stable mass transfer binaries from \citet{Rappaport_1995}, with the dotted line representing an update on this from \citet{Lin_2011}. Systems above this region may be explained by non-conservative stable mass transfer.}
    \label{fig:PMass}
\end{figure*}

In Figure~\ref{fig:PMass}, we show the orbital period vs the white dwarf mass distribution, with two panels - one each for the more robust astrometric data, and for the lower mass estimates from the spectroscopic data. As can be observed, we have a wide range of orbital periods from our candidate systems. We can see that for our astrometric binaries (top panel) our systems lie near the region for post-stable mass transfer systems as proposed by \citet{Rappaport_1995} (the grey shaded region), with few outliers. A similar result was recently presented by \citet{Shahaf_2023b} using a purely astrometrically-selected sample of systems. We note that this relation is based on donor stars on the Red Giant Branch (RGB), which appears to be the case for many of our systems where the WD mass is $<0.5$\,M$_\odot$, but this evolutionary channel may not be able to explain the origin of systems with higher mass WDs that likely reached the Asymptotic Giant Branch (AGB). A small but significant number of our systems sit just below the stable mass transfer region, with similar parameters to the self-lensing systems discovered by the \textit{Kepler} mission \citep{Kruse_2014,Kawahara_2018}. Given their long periods it is still challenging to explain them as a result of common envelope evolution, even assuming very efficient envelope removal. Interestingly, the luminous star in many of these systems is somewhat evolved itself, generally on the subgiant branch. This is unsurprising if these systems are the result of stable mass transfer, since this requires an initial mass ratio close to unity and therefore similar main-sequence lifetimes for the two components. Given their current stellar and binary parameters these systems are likely to undergo a common envelope phase at some point in the future. Should both components survive this additional mass transfer phase, this would lead to the creation of a double WD binary, and could represent a significant formation channel for such systems.

The spectroscopically identified systems (bottom panel of Figure~\ref{fig:PMass}) show much wider distributions for both period and WD mass. Given that we only have lower limits on the WD masses the picture is less clear for these systems. There are some clear post-common envelope systems with periods less than a couple of weeks and minimum WD masses around 0.5~M\textsubscript{$\odot$}, similar to other systems found in the white dwarf binary pathways survey. However, there is a significant number of systems with very low minimum WD masses, covering a wide range of periods. While these could all be low inclination post-common envelope systems, there is also a possibility that there is a population of post-stable mass transfer systems ranging from periods of a day up to 1000 days. Indeed the shortest periods systems, EL~CVn binaries, are already known and intermediate period post-stable mass transfer binaries have now also been found \citep{PathwayIX}. Given the low WD masses, these may well be the progenitors of double WD systems containing at least one extremely low mass WD. 

We also note that there are a number of systems in both the astrometric and spectroscopic samples with clearly high mass WDs and orbital periods longer than a few weeks. The high WD masses make these systems challenging to explain as the result of stable mass transfer, but equally their long orbital periods are challenging to explain via standard common envelope evolution. Recently \citet{Yamaguchi_2023} discovered 5 such systems occupying this parameter space (albeit with slightly shorter orbital periods than many of our systems) which, combined with the roughly 25 such systems we find, clearly indicates that there is a population of WD binaries in this region of parameter space. \citet{Yamaguchi_2023} proposed that these systems could be the result of common envelope evolution if internal (recombination) energy also aids the expulsion of the envelope. Whether this can also explain the systems we find with orbital periods of 100s of days is unclear.

One clear result appears to be that a far wider range of periods is viable for WD\,+\,FGK systems compared to WD\,+\,dM binaries, which generally have periods averaging around 10.3~hours {\citep{Nebot_2011}}, with a very small sample of WD\,+\,dM binaries from \citet{Shahaf_2023b} having relatively long periods being the current major exception, whilst WD\,+\,FGK systems can span periods from around half a day to almost 1000~days. This is likely a result of the multiple different formation pathways possible for WD\,+\,FGK systems, whereas the extreme initial mass ratios of WD\,+\,dM systems mean that they are all likely the result of standard common envelope evolution.

\subsection{Systems with candidate hot subdwarf stars}
\label{sec:sdOB}

We made a somewhat empirical cut to determine if the system was a hot subdwarf candidate, by flagging systems with an absolute $FUV$ magnitude brighter than 5.2\,mag as being hot subdwarf binary candidate systems, opposed to WD binary systems. Doing so, we have found three systems in our sample that are candidate hot subdwarf binaries, as can be observed in Figure~\ref{fig:Optical_HRs}, given their extreme UV luminosities. Hot subdwarf stars are core helium burning stars that have been stripped of their envelope \citep{Herber_2009}, with masses in the domain of $\sim$\,0.4-0.5\,M\textsubscript{$\odot$} \citep{Vos_2018}, and are bright in the ultraviolet, similar to WDs - though they are brighter at optically blue wavelengths in comparison. This means that the cut to remove systems a magnitude bluer than the main sequence track made in \citet{PathwayI}, \citet{PathwayII}, \citet{PathwayV} and our present study should remove such systems in a binary with a dwarf late F, G or K companion. However, such a cut would not remove hot subdwarfs with a more evolved, hence luminous companion, as a similar cut was not made along the giant branch. Hot subdwarf binaries typically come in two groups; those with low mass companions in short period systems (such as those discussed in \citealt{Kupfer_2015}), and those with higher mass companions and a longer orbital period (as discussed in \citealt{Vos_2018}).

Our three hot subdwarf candidates are in binaries with evolved companions (see Figure~\ref{fig:Optical_HRs}), and are found spectroscopically. Two of our three systems, UCAC2~15655859 and TYC~4889-1238-1, possess periods that are consistent with the established period domain of hot subdwarfs with FGK stars from \citet{Vos_2018} (having periods of $\sim$\,996 and 1079 days respectively, and minimum masses of $\sim$\,0.19\,M\textsubscript{$\odot$}). The third candidate, TYC~6472-1267-1, has a period of around 56 days, which is unusual, as there is a dearth of hot subdwarf binary systems with such periods (ignoring the systems with WD companions, which have passed through two mass transfer phases and so are not directly comparable). It is worth noting that this system is a candidate triple system based on the findings of \citet{Tian_2020}, who flagged the system as potentially having a common proper motion companion. If true (which is somewhat dubious given the separation of 55302~au found by \citealt{Tian_2020}), then this would be consistent with the findings of \citet{Lagos_2020} and \citet{PathwayIX}, that these intermediate period post-stable mass transfer systems are generally found in triples, since they need to have very short initial separations. We do note however, that extremely low mass WDs can also have similar UV luminosities to hot subdwarf stars, so, while these are clearly not standard mass WDs, they may not be hot subdwarf stars either and follow up data are required to properly classify these objects.

\subsection{UV Parameter Space}

\begin{figure*}
    \centering
    \includegraphics[height=0.9\textheight]{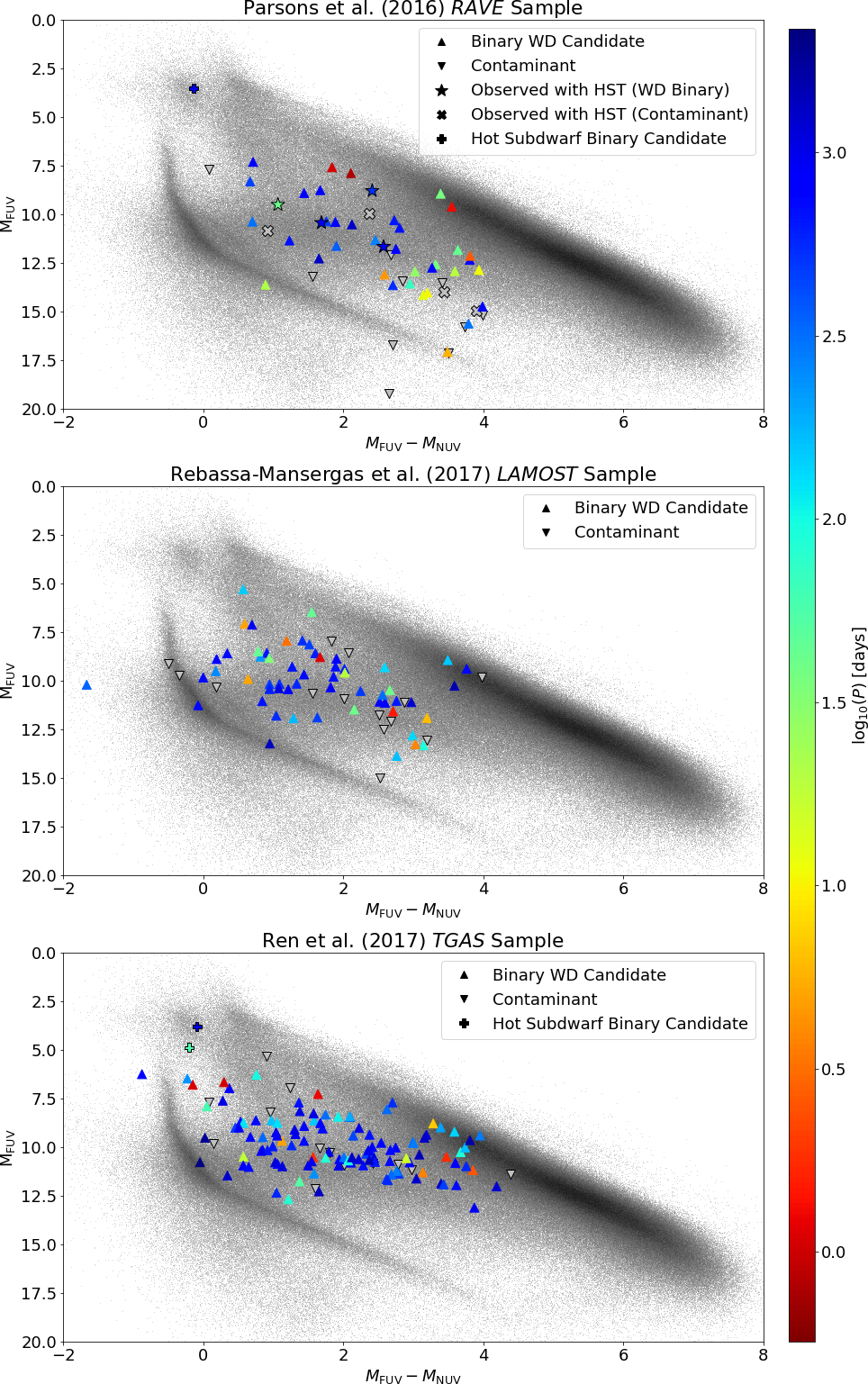}
    \caption{UV colour-magnitude diagrams using \textit{GALEX} $FUV$ and $NUV$ magnitudes, split by the three surveys of \citet{PathwayI}, \citet{PathwayII} and \citet{PathwayV}. The points are colour-coded by the log\textsubscript{10} of their period. A random sample of stars within 250~pc are shown in the background.}
    \label{fig:UV_HRs}
\end{figure*}

Given that both the \textit{RAVE} and \textit{LAMOST} samples were created before any \textit{Gaia} data releases, it is useful to see where these systems sit in a UV colour-magnitude diagram, both the genuine WD\,+\,FGK systems as well as the contaminants. This may aid future selections of these kinds of objects. Figure~\ref{fig:UV_HRs} shows such a UV colour-magnitude diagram, separated by the systems found by \citet{PathwayI}, \citet{PathwayII} and \citet{PathwayV}.

Most systems sit between the WD and main-sequence track as expected for these kinds of objects. In general, there is more contamination towards fainter objects, with almost all sources fainter than $M_{FUV}\sim15$\,mag seen to be contaminants, likely active stars. However, there are also brighter contaminants that sit right in the main population of WD\,+\,FGK systems, including some that we have observed with \textit{HST}, confirming that they are indeed contaminants. In general, these tend to be triple systems where the WD is a wide companion to a main-sequence binary. We also note that there appears to be some genuine WD\,+\,FGK systems that show very minor UV excesses (those that sit close to the main-sequence), which are likely systems containing cool and/or high mass WDs. However, there is also significant contamination close to the main-sequence, so caution is needed when searching for WD binaries in this region.

We note that the largest source of contamination is generally due to active stars, i.e. binaries consisting of an FGK star plus a lower mass companion, where either one or both components have strong chromospheric emission in the UV. Despite this, we advise against removing active stars from catalogues before searching for UV excess systems, because many genuine WD\,+\,FGK systems contain an active FGK star, since they are often spun up as a result of mass transfer and/or tidal locking.

\subsection{Astrometric Parameter Space}
\label{sec:AstroSpace}

\begin{figure}
    \centering
    \includegraphics[width=\columnwidth]{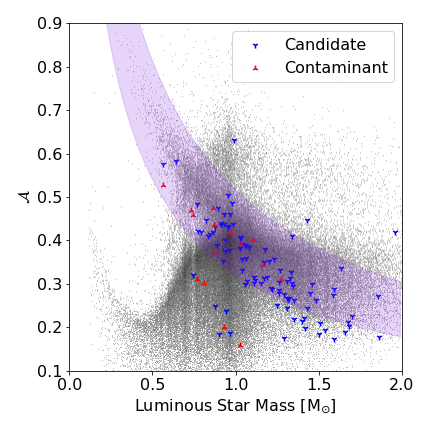}
    \caption{The relationship between the astrometric mass function of \citet{Shahaf_2019}, $\mathcal{A}$, and the mass of the luminous companion. The purple highlighted region represents the region where the hidden companions mass is between 0.4-0.75~M\textsubscript{$\odot$}, a typical mass range for WDs in these systems. The full sample of astrometric binaries in \textit{Gaia} is shown in the background.}
    \label{fig:Astromass}
\end{figure}

It is now possible to identify WD\,+\,FGK systems directly as astrometric binaries without the need for any UV data, by means of their astrometric mass function, $\mathcal{A}$, as described in \citet{Shahaf_2019}, and removing systems likely to be either a MS~+~MS or a hierarchical triple system. \citet{Shahaf_2023b} recently used this technique to identify several thousand candidate WD\,+\,FGK systems in \textit{Gaia}. Given that our selection criteria are significantly different to theirs it is worth comparing the two methods. In particular, in Figure~\ref{fig:Astromass} we plot the astrometric mass function for our sample and indicate the typical region that a WD\,+\,FGK system is likely to reside. The  \citet{Shahaf_2023b} method relies on ruling out the possibility of a binary being composed of either two luminous stars, or the companion being a binary itself (i.e. a triple system), which generally requires placing a fairly tight limit of $\mathcal{A}$ greater than around 0.4 (see \citealt{Shahaf_2023b} for the exact cut). Applying the same cut to our samples would exclude a significant fraction of all our systems. This is due to a combination of many of our WD masses being significantly lower than the canonical WD mass (possibly as a result of stable mass transfer) and our sample containing a large number of more massive luminous stars, since the \citet{Shahaf_2023b} sample cuts off at around 1.3~M\textsubscript{$\odot$}, where contamination from normal binaries overwhelms the WD systems. Many of our luminous stars are also giants, which were not included in the \citet{Shahaf_2023b} sample. Therefore, the UV-excess method appears to be the only way to identify systems with very low mass WDs and/or higher mass luminous stars at present. We do note that some of our systems sit quite outside the region expected for WD companions (i.e. the purple region in Figure~\ref{fig:Astromass}) and while some of these objects have been identified as contaminants, it is possible that our candidates significantly above and below the WD companion region may also no be genuine WD\,+\,FGK systems.

As one could imagine, our sample is bias towards systems with a large UV excess - namely hotter, lower mass WDs, whilst our cut removing systems bluer than the main sequence as discussed in Section~\ref{sec:Targets} is bias against lower mass luminous companions where the WD may contribute a significant percentage of the optical flux. There are further more complex selection effects in each of the parent samples contributing to our sample, see \citet{PathwayI}, \citet{PathwayII} and \citet{PathwayV} for more information on each. However, our bias towards hot, low mass WDs is complimentary to the selection effects of the purely astrometric method of \citet{Shahaf_2023b}, which is somewhat bias towards higher mass WDs, so a combination of both methods could lead to a more complete sample.

\section{Conclusions}
\label{sec:Conclusion}

We have confirmed the binary nature of 246 candidate WD\,+\,FGK binaries from \citet{PathwayI}, \citet{PathwayII} and \citet{PathwayV} using data from \textit{Gaia} and combined this with modelling the SED of the luminous star in order to constrain the mass of the unseen WD. Once systems which have been flagged as likely contaminants are set aside, we have a sample of 206 WD\,+\,FGK binary candidates, four of which have been confirmed through \textit{HST} spectroscopic observations in the ultraviolet and one of which has been published before \citep{PathwayIX}.

We find that our astrometric systems occupy a region of parameter space that is incompatible with a post-CE origin, given their low WD masses, but they do lie near the relationship proposed by \citet{Rappaport_1995} for post-stable mass transfer binaries, indicating these systems likely underwent a period of stable, non-conservative mass transfer earlier in their evolution. Many of these systems are likely to be the progenitors of double WD binaries.

The analysis of the spectroscopically  identified systems is complicated by the fact that we can only set a lower limit on the WD parameters. Nevertheless, they confirm the existence of WD+FGK binaries at a wide range of orbital periods, from less than a day, to more than 1000 days, in contrast to the population of WD\,+\,dM binaries. We also identify a group of systems with orbital periods of more than a few weeks containing high mass WDs, which are difficult to reproduce either via stable mass transfer or standard $\alpha_\mathrm{CE}\sim0.3$ common envelope evolution.

\section*{Acknowledgements}

This work has made use of data from the European Space Agency (ESA) mission {\it Gaia} (\url{https://www.cosmos.esa.int/gaia}), processed by the {\it Gaia} Data Processing and Analysis Consortium (DPAC, \url{https://www.cosmos.esa.int/web/gaia/dpac/consortium}). Funding for the DPAC has been provided by national institutions, in particular the institutions participating in the {\it Gaia} Multilateral Agreement. SGP acknowledges the support of a Science and Technology Facilities Council (STFC) Ernest Rutherford fellowship. This work has been partially supported by the Spanish MINECO grant PID2020-117252GB-I00 and by the AGAUR/Generalitat de Catalunya grant SGR-386/2021. RR acknowledges support from Grant RYC2021-030837-I funded by MCIN/AEI/ 10.13039/501100011033 and by “European Union NextGeneration EU/PRTR”. MZ and MRS acknowledge support from Fondecyt (grant 1221059). This project has received funding from the European Research Council (ERC) under the European Union's Horizon 2020 research and innovation programme (Grant agreement No. 101020057). For the purpose of open access, the author has applied a creative commons attribution (CC BY) licence to any author accepted manuscript version arising. 

\section*{Data Availability}

The previous data samples which this paper built on are upon are available publicly online. Likewise \textit{Gaia} data release 3 is available publicly online at \url{https://www.cosmos.esa.int/web/gaia/dr3}. Raw and reduced \textit{Hubble} data are available through MAST at \url{https://mast.stsci.edu/portal/Mashup/Clients/Mast/Portal.html}



\bibliographystyle{mnras}
\bibliography{Bib} 

\begin{thebibliography}{}
\makeatletter
\relax
\def\mn@urlcharsother{\let\do\@makeother \do\$\do\&\do\#\do\^\do\_\do\%\do\~}
\def\mn@doi{\begingroup\mn@urlcharsother \@ifnextchar [ {\mn@doi@}
  {\mn@doi@[]}}
\def\mn@doi@[#1]#2{\def\@tempa{#1}\ifx\@tempa\@empty \href
  {http://dx.doi.org/#2} {doi:#2}\else \href {http://dx.doi.org/#2} {#1}\fi
  \endgroup}
\def\mn@eprint#1#2{\mn@eprint@#1:#2::\@nil}
\def\mn@eprint@arXiv#1{\href {http://arxiv.org/abs/#1} {{\tt arXiv:#1}}}
\def\mn@eprint@dblp#1{\href {http://dblp.uni-trier.de/rec/bibtex/#1.xml}
  {dblp:#1}}
\def\mn@eprint@#1:#2:#3:#4\@nil{\def\@tempa {#1}\def\@tempb {#2}\def\@tempc
  {#3}\ifx \@tempc \@empty \let \@tempc \@tempb \let \@tempb \@tempa \fi \ifx
  \@tempb \@empty \def\@tempb {arXiv}\fi \@ifundefined
  {mn@eprint@\@tempb}{\@tempb:\@tempc}{\expandafter \expandafter \csname
  mn@eprint@\@tempb\endcsname \expandafter{\@tempc}}}

\bibitem[\protect\citeauthoryear{{Althaus}, {Miller Bertolami}  \&
  {C{\'o}rsico}}{{Althaus} et~al.}{2013}]{Althaus_2013}
{Althaus} L.~G.,  {Miller Bertolami} M.~M.,   {C{\'o}rsico} A.~H.,  2013,
  \mn@doi [\aap] {10.1051/0004-6361/201321868}, \href
  {https://ui.adsabs.harvard.edu/abs/2013A&A...557A..19A} {557, A19}

\bibitem[\protect\citeauthoryear{{B{\'e}dard}, {Bergeron}, {Brassard}  \&
  {Fontaine}}{{B{\'e}dard} et~al.}{2020}]{Bedard2020}
{B{\'e}dard} A.,  {Bergeron} P.,  {Brassard} P.,   {Fontaine} G.,  2020,
  \mn@doi [\apj] {10.3847/1538-4357/abafbe}, \href
  {https://ui.adsabs.harvard.edu/abs/2020ApJ...901...93B} {901, 93}

\bibitem[\protect\citeauthoryear{{Belloni}, {Zorotovic}, {Schreiber}, {Parsons}
   \& {Garbutt}}{{Belloni} et~al.}{2024}]{Belloni_2024}
{Belloni} D.,  {Zorotovic} M.,  {Schreiber} M.~R.,  {Parsons} S.~G.,
  {Garbutt} J.~A.,  2024, \mn@doi [arXiv e-prints] {10.48550/arXiv.2401.07692},
  \href {https://ui.adsabs.harvard.edu/abs/2024arXiv240107692B} {p.
  arXiv:2401.07692}

\bibitem[\protect\citeauthoryear{Bédard, Bergeron, Brassard  \&
  Fontaine}{Bédard et~al.}{2020}]{Bédard_2020}
Bédard A.,  Bergeron P.,  Brassard P.,   Fontaine G.,  2020, \mn@doi [The
  Astrophysical Journal] {10.3847/1538-4357/abafbe}, 901, 93

\bibitem[\protect\citeauthoryear{{Capitanio}, {Lallement}, {Vergely},
  {Elyajouri}  \& {Monreal-Ibero}}{{Capitanio} et~al.}{2017}]{Stilism}
{Capitanio} L.,  {Lallement} R.,  {Vergely} J.~L.,  {Elyajouri} M.,
  {Monreal-Ibero} A.,  2017, \mn@doi [\aap] {10.1051/0004-6361/201730831},
  \href {https://ui.adsabs.harvard.edu/abs/2017A&A...606A..65C} {606, A65}

\bibitem[\protect\citeauthoryear{{Casagrande}, {Sch{\"o}nrich}, {Asplund},
  {Cassisi}, {Ram{\'\i}rez}, {Mel{\'e}ndez}, {Bensby}  \&
  {Feltzing}}{{Casagrande} et~al.}{2011}]{Casagrande_2011}
{Casagrande} L.,  {Sch{\"o}nrich} R.,  {Asplund} M.,  {Cassisi} S.,
  {Ram{\'\i}rez} I.,  {Mel{\'e}ndez} J.,  {Bensby} T.,   {Feltzing} S.,  2011,
  \mn@doi [\aap] {10.1051/0004-6361/201016276}, \href
  {https://ui.adsabs.harvard.edu/abs/2011A&A...530A.138C} {530, A138}

\bibitem[\protect\citeauthoryear{{Choi}, {Dotter}, {Conroy}, {Cantiello},
  {Paxton}  \& {Johnson}}{{Choi} et~al.}{2016}]{MIST1}
{Choi} J.,  {Dotter} A.,  {Conroy} C.,  {Cantiello} M.,  {Paxton} B.,
  {Johnson} B.~D.,  2016, \mn@doi [\apj] {10.3847/0004-637X/823/2/102}, \href
  {https://ui.adsabs.harvard.edu/abs/2016ApJ...823..102C} {823, 102}

\bibitem[\protect\citeauthoryear{Cummings, Kalirai, Tremblay, Ramirez-Ruiz  \&
  Choi}{Cummings et~al.}{2018}]{Cummings_2018}
Cummings J.~D.,  Kalirai J.~S.,  Tremblay P.-E.,  Ramirez-Ruiz E.,   Choi J.,
  2018, \mn@doi [The Astrophysical Journal] {10.3847/1538-4357/aadfd6}, 866, 21

\bibitem[\protect\citeauthoryear{{Cutri} et~al.,}{{Cutri}
  et~al.}{2021}]{AllWISE}
{Cutri} R.~M.,  et~al., 2021, VizieR Online Data Catalog, \href
  {https://ui.adsabs.harvard.edu/abs/2014yCat.2328....0C} {p. II/328}

\bibitem[\protect\citeauthoryear{{Dotter}}{{Dotter}}{2016}]{MIST0}
{Dotter} A.,  2016, \mn@doi [\apjs] {10.3847/0067-0049/222/1/8}, \href
  {https://ui.adsabs.harvard.edu/abs/2016ApJS..222....8D} {222, 8}

\bibitem[\protect\citeauthoryear{{Foreman-Mackey}, {Hogg}, {Lang}  \&
  {Goodman}}{{Foreman-Mackey} et~al.}{2013}]{Foreman2013}
{Foreman-Mackey} D.,  {Hogg} D.~W.,  {Lang} D.,   {Goodman} J.,  2013, \mn@doi
  [\pasp] {10.1086/670067}, \href
  {https://ui.adsabs.harvard.edu/abs/2013PASP..125..306F} {125, 306}

\bibitem[\protect\citeauthoryear{{Gaia Collaboration} et~al.,}{{Gaia
  Collaboration} et~al.}{2022}]{Gaia_Stellar_Multiplicity}
{Gaia Collaboration} et~al., 2022, \mn@doi [arXiv e-prints]
  {10.48550/arXiv.2206.05595}, \href
  {https://ui.adsabs.harvard.edu/abs/2022arXiv220605595G} {p. arXiv:2206.05595}

\bibitem[\protect\citeauthoryear{{Halbwachs} et~al.,}{{Halbwachs}
  et~al.}{2022}]{Gaia_Processing}
{Halbwachs} J.-L.,  et~al., 2022, \mn@doi [arXiv e-prints]
  {10.48550/arXiv.2206.05726}, \href
  {https://ui.adsabs.harvard.edu/abs/2022arXiv220605726H} {p. arXiv:2206.05726}

\bibitem[\protect\citeauthoryear{{Heber}}{{Heber}}{2009}]{Herber_2009}
{Heber} U.,  2009, \mn@doi [\araa] {10.1146/annurev-astro-082708-101836}, \href
  {https://ui.adsabs.harvard.edu/abs/2009ARA&A..47..211H} {47, 211}

\bibitem[\protect\citeauthoryear{Hernandez et~al.,}{Hernandez
  et~al.}{2020}]{PathwayIV}
Hernandez M.~S.,  et~al., 2020, \mn@doi [Monthly Notices of the Royal
  Astronomical Society] {10.1093/mnras/staa3815}, 501, 1677

\bibitem[\protect\citeauthoryear{Hernandez et~al.,}{Hernandez
  et~al.}{2022a}]{PathwayVI}
Hernandez M.~S.,  et~al., 2022a, \mn@doi [Monthly Notices of the Royal
  Astronomical Society] {10.1093/mnras/stac604}, 512, 1843

\bibitem[\protect\citeauthoryear{Hernandez et~al.,}{Hernandez
  et~al.}{2022b}]{PathwayVIII}
Hernandez M.~S.,  et~al., 2022b, \mn@doi [Monthly Notices of the Royal
  Astronomical Society] {10.1093/mnras/stac2837}, 517, 2867

\bibitem[\protect\citeauthoryear{Holberg}{Holberg}{2009}]{Holberg_2009}
Holberg J.~B.,  2009, \mn@doi [Journal of Physics: Conference Series]
  {10.1088/1742-6596/172/1/012022}, 172, 012022

\bibitem[\protect\citeauthoryear{Kawahara, Masuda, MacLeod, Latham, Bieryla  \&
  Benomar}{Kawahara et~al.}{2018}]{Kawahara_2018}
Kawahara H.,  Masuda K.,  MacLeod M.,  Latham D.~W.,  Bieryla A.,   Benomar O.,
   2018, \mn@doi [The Astronomical Journal] {10.3847/1538-3881/aaaaaf}, 155,
  144

\bibitem[\protect\citeauthoryear{Kepler, Kleinman, Nitta, Koester, Castanheira,
  Giovannini, Costa  \& Althaus}{Kepler et~al.}{2007}]{Kepler_2007}
Kepler S.~O.,  Kleinman S.~J.,  Nitta A.,  Koester D.,  Castanheira B.~G.,
  Giovannini O.,  Costa A. F.~M.,   Althaus L.,  2007, \mn@doi [Monthly Notices
  of the Royal Astronomical Society] {10.1111/j.1365-2966.2006.11388.x}, 375,
  1315

\bibitem[\protect\citeauthoryear{{Koester}}{{Koester}}{2010}]{Koester2010}
{Koester} D.,  2010, \memsai, \href
  {https://ui.adsabs.harvard.edu/abs/2010MmSAI..81..921K} {81, 921}

\bibitem[\protect\citeauthoryear{Kruse \& Agol}{Kruse \&
  Agol}{2014a}]{KruseAgol_2014}
Kruse E.,  Agol E.,  2014a, \mn@doi [Science] {10.1126/science.1251999}, 344,
  275

\bibitem[\protect\citeauthoryear{{Kruse} \& {Agol}}{{Kruse} \&
  {Agol}}{2014b}]{Kruse_2014}
{Kruse} E.,  {Agol} E.,  2014b, \mn@doi [Science] {10.1126/science.1251999},
  \href {https://ui.adsabs.harvard.edu/abs/2014Sci...344..275K} {344, 275}

\bibitem[\protect\citeauthoryear{{Kupfer} et~al.,}{{Kupfer}
  et~al.}{2015}]{Kupfer_2015}
{Kupfer} T.,  et~al., 2015, \mn@doi [A\&A] {10.1051/0004-6361/201425213}, 576,
  A44

\bibitem[\protect\citeauthoryear{{Lagos} et~al.,}{{Lagos}
  et~al.}{2020a}]{PathwayIII}
{Lagos} F.,  et~al., 2020a, \mn@doi [\mnras] {10.1093/mnras/staa747}, \href
  {https://ui.adsabs.harvard.edu/abs/2020MNRAS.494..915L} {494, 915}

\bibitem[\protect\citeauthoryear{Lagos, Schreiber, Parsons, Gänsicke  \&
  Godoy}{Lagos et~al.}{2020b}]{Lagos_2020}
Lagos F.,  Schreiber M.~R.,  Parsons S.~G.,  Gänsicke B.~T.,   Godoy N.,
  2020b, \mn@doi [Monthly Notices of the Royal Astronomical Society: Letters]
  {10.1093/mnrasl/slaa164}, 499, L121

\bibitem[\protect\citeauthoryear{{Lagos}, {Schreiber}, {Parsons}, {Toloza},
  {G{\"a}nsicke}, {Hernandez}, {Schmidtobreick}  \& {Belloni}}{{Lagos}
  et~al.}{2022}]{PathwayVII}
{Lagos} F.,  {Schreiber} M.~R.,  {Parsons} S.~G.,  {Toloza} O.,  {G{\"a}nsicke}
  B.~T.,  {Hernandez} M.~S.,  {Schmidtobreick} L.,   {Belloni} D.,  2022,
  \mn@doi [\mnras] {10.1093/mnras/stac673}, \href
  {https://ui.adsabs.harvard.edu/abs/2022MNRAS.512.2625L} {512, 2625}

\bibitem[\protect\citeauthoryear{Lin, Rappaport, Podsiadlowski, Nelson, Paxton
  \& Todorov}{Lin et~al.}{2011}]{Lin_2011}
Lin J.,  Rappaport S.,  Podsiadlowski P.,  Nelson L.,  Paxton B.,   Todorov P.,
   2011, \mn@doi [The Astrophysical Journal] {10.1088/0004-637X/732/2/70}, 732,
  70

\bibitem[\protect\citeauthoryear{{Martin} et~al.,}{{Martin}
  et~al.}{2005}]{GALEX}
{Martin} D.~C.,  et~al., 2005, \mn@doi [\apjl] {10.1086/426387}, \href
  {https://ui.adsabs.harvard.edu/abs/2005ApJ...619L...1M} {619, L1}

\bibitem[\protect\citeauthoryear{Masuda, Kawahara, Latham, Bieryla, Kunitomo,
  MacLeod  \& Aoki}{Masuda et~al.}{2019}]{Masuda_2019}
Masuda K.,  Kawahara H.,  Latham D.~W.,  Bieryla A.,  Kunitomo M.,  MacLeod M.,
    Aoki W.,  2019, \mn@doi [The Astrophysical Journal Letters]
  {10.3847/2041-8213/ab321b}, 881, L3

\bibitem[\protect\citeauthoryear{{Moreno}, {Schneider}, {R{\"o}pke}, {Ohlmann},
  {Pakmor}, {Podsiadlowski}  \& {Sand}}{{Moreno} et~al.}{2022}]{Moreno_2022}
{Moreno} M.~M.,  {Schneider} F. R.~N.,  {R{\"o}pke} F.~K.,  {Ohlmann} S.~T.,
  {Pakmor} R.,  {Podsiadlowski} P.,   {Sand} C.,  2022, \mn@doi [\aap]
  {10.1051/0004-6361/202142731}, \href
  {https://ui.adsabs.harvard.edu/abs/2022A&A...667A..72M} {667, A72}

\bibitem[\protect\citeauthoryear{{Morton}}{{Morton}}{2015}]{Isochrones_Package}
{Morton} T.~D.,  2015, {isochrones: Stellar model grid package}, Astrophysics
  Source Code Library, record ascl:1503.010 (\mn@eprint {ascl} {1503.010})

\bibitem[\protect\citeauthoryear{{Nebot G{\'o}mez-Mor{\'a}n} et~al.,}{{Nebot
  G{\'o}mez-Mor{\'a}n} et~al.}{2011}]{Nebot_2011}
{Nebot G{\'o}mez-Mor{\'a}n} A.,  et~al., 2011, \mn@doi [\aap]
  {10.1051/0004-6361/201117514}, \href
  {https://ui.adsabs.harvard.edu/abs/2011A&A...536A..43N} {536, A43}

\bibitem[\protect\citeauthoryear{{Nelemans}, {Verbunt}, {Yungelson}  \&
  {Portegies Zwart}}{{Nelemans} et~al.}{2000}]{Nelemans_2000}
{Nelemans} G.,  {Verbunt} F.,  {Yungelson} L.~R.,   {Portegies Zwart} S.~F.,
  2000, \mn@doi [\aap] {10.48550/arXiv.astro-ph/0006216}, \href
  {https://ui.adsabs.harvard.edu/abs/2000A&A...360.1011N} {360, 1011}

\bibitem[\protect\citeauthoryear{{Ohlmann}, {R{\"o}pke}, {Pakmor}  \&
  {Springel}}{{Ohlmann} et~al.}{2016}]{Ohlmann_2016}
{Ohlmann} S.~T.,  {R{\"o}pke} F.~K.,  {Pakmor} R.,   {Springel} V.,  2016,
  \mn@doi [\apjl] {10.3847/2041-8205/816/1/L9}, \href
  {https://ui.adsabs.harvard.edu/abs/2016ApJ...816L...9O} {816, L9}

\bibitem[\protect\citeauthoryear{{Ondratschek, Patrick A.}, {R\"opke, Friedrich
  K.}, {Schneider, Fabian R. N.}, {Fendt, Christian}, {Sand, Christian},
  {Ohlmann, Sebastian T.}, {Pakmor, R\"udiger}  \& {Springel,
  Volker}}{{Ondratschek, Patrick A.} et~al.}{2022}]{Ondratschek_2022}
{Ondratschek, Patrick A.} {R\"opke, Friedrich K.} {Schneider, Fabian R. N.}
  {Fendt, Christian} {Sand, Christian} {Ohlmann, Sebastian T.} {Pakmor,
  R\"udiger}  {Springel, Volker} 2022, \mn@doi [A\&A]
  {10.1051/0004-6361/202142478}, 660, L8

\bibitem[\protect\citeauthoryear{{Paczynski}}{{Paczynski}}{1976}]{Paczynski_1976}
{Paczynski} B.,  1976, in {Eggleton} P.,  {Mitton} S.,   {Whelan} J.,  eds,
  Proceedings of the Symposium Vol. 73, Structure and Evolution of Close Binary
  Systems. p.~75

\bibitem[\protect\citeauthoryear{Parsons et~al.,}{Parsons
  et~al.}{2015}]{Parsons_2015}
Parsons S.~G.,  et~al., 2015, \mn@doi [Monthly Notices of the Royal
  Astronomical Society] {10.1093/mnras/stv1395}, 452, 1754

\bibitem[\protect\citeauthoryear{{Parsons}, {Rebassa-Mansergas}, {Schreiber},
  {G{\"a}nsicke}, {Zorotovic}  \& {Ren}}{{Parsons} et~al.}{2016}]{PathwayI}
{Parsons} S.~G.,  {Rebassa-Mansergas} A.,  {Schreiber} M.~R.,  {G{\"a}nsicke}
  B.~T.,  {Zorotovic} M.,   {Ren} J.~J.,  2016, \mn@doi [\mnras]
  {10.1093/mnras/stw2143}, \href
  {https://ui.adsabs.harvard.edu/abs/2016MNRAS.463.2125P} {463, 2125}

\bibitem[\protect\citeauthoryear{{Parsons} et~al.,}{{Parsons}
  et~al.}{2023}]{PathwayIX}
{Parsons} S.~G.,  et~al., 2023, \mn@doi [\mnras] {10.1093/mnras/stac3368},
  \href {https://ui.adsabs.harvard.edu/abs/2023MNRAS.518.4579P} {518, 4579}

\bibitem[\protect\citeauthoryear{{Passy} et~al.,}{{Passy}
  et~al.}{2012}]{Passy_2012}
{Passy} J.-C.,  et~al., 2012, \mn@doi [\apj] {10.1088/0004-637X/744/1/52},
  \href {https://ui.adsabs.harvard.edu/abs/2012ApJ...744...52P} {744, 52}

\bibitem[\protect\citeauthoryear{{Paxton}, {Bildsten}, {Dotter}, {Herwig},
  {Lesaffre}  \& {Timmes}}{{Paxton} et~al.}{2011}]{MESA_Stel_Astro}
{Paxton} B.,  {Bildsten} L.,  {Dotter} A.,  {Herwig} F.,  {Lesaffre} P.,
  {Timmes} F.,  2011, \mn@doi [\apjs] {10.1088/0067-0049/192/1/3}, \href
  {https://ui.adsabs.harvard.edu/abs/2011ApJS..192....3P} {192, 3}

\bibitem[\protect\citeauthoryear{{Paxton} et~al.,}{{Paxton}
  et~al.}{2013}]{MESA_PORMS}
{Paxton} B.,  et~al., 2013, \mn@doi [\apjs] {10.1088/0067-0049/208/1/4}, \href
  {https://ui.adsabs.harvard.edu/abs/2013ApJS..208....4P} {208, 4}

\bibitem[\protect\citeauthoryear{{Paxton} et~al.,}{{Paxton}
  et~al.}{2015}]{MESA_BPE}
{Paxton} B.,  et~al., 2015, \mn@doi [\apjs] {10.1088/0067-0049/220/1/15}, \href
  {https://ui.adsabs.harvard.edu/abs/2015ApJS..220...15P} {220, 15}

\bibitem[\protect\citeauthoryear{Podsiadlowski}{Podsiadlowski}{2014}]{Podsiadlowski_2014}
Podsiadlowski P.,  2014, Accretion Processes in Astrophysics, p.~45

\bibitem[\protect\citeauthoryear{Rappaport, Podsiadlowski, Joss, Di~Stefano  \&
  Han}{Rappaport et~al.}{1995}]{Rappaport_1995}
Rappaport S.,  Podsiadlowski P.,  Joss P.~C.,  Di~Stefano R.,   Han Z.,  1995,
  \mn@doi [Monthly Notices of the Royal Astronomical Society]
  {10.1093/mnras/273.3.731}, 273, 731

\bibitem[\protect\citeauthoryear{Rebassa-Mansergas et~al.,}{Rebassa-Mansergas
  et~al.}{2008}]{Rebassa-Mansergas_2008}
Rebassa-Mansergas A.,  et~al., 2008, \mn@doi [Monthly Notices of the Royal
  Astronomical Society] {10.1111/j.1365-2966.2008.13850.x}, 390, 1635

\bibitem[\protect\citeauthoryear{{Rebassa-Mansergas}
  et~al.,}{{Rebassa-Mansergas} et~al.}{2017}]{PathwayII}
{Rebassa-Mansergas} A.,  et~al., 2017, \mn@doi [\mnras]
  {10.1093/mnras/stx2259}, \href
  {https://ui.adsabs.harvard.edu/abs/2017MNRAS.472.4193R} {472, 4193}

\bibitem[\protect\citeauthoryear{{Ren} et~al.,}{{Ren} et~al.}{2020}]{PathwayV}
{Ren} J.~J.,  et~al., 2020, \mn@doi [\apj] {10.3847/1538-4357/abc017}, \href
  {https://ui.adsabs.harvard.edu/abs/2020ApJ...905...38R} {905, 38}

\bibitem[\protect\citeauthoryear{{Riello} et~al.,}{{Riello}
  et~al.}{2021}]{Gaia_Photometric}
{Riello} M.,  et~al., 2021, \mn@doi [\aap] {10.1051/0004-6361/202039587}, \href
  {https://ui.adsabs.harvard.edu/abs/2021A&A...649A...3R} {649, A3}

\bibitem[\protect\citeauthoryear{{Scherbak} \& {Fuller}}{{Scherbak} \&
  {Fuller}}{2023}]{Scherbak_2023}
{Scherbak} P.,  {Fuller} J.,  2023, \mn@doi [\mnras] {10.1093/mnras/stac3313},
  \href {https://ui.adsabs.harvard.edu/abs/2023MNRAS.518.3966S} {518, 3966}

\bibitem[\protect\citeauthoryear{{Shahaf}, {Mazeh}, {Faigler}  \&
  {Holl}}{{Shahaf} et~al.}{2019}]{Shahaf_2019}
{Shahaf} S.,  {Mazeh} T.,  {Faigler} S.,   {Holl} B.,  2019, \mn@doi [\mnras]
  {10.1093/mnras/stz1636}, \href
  {https://ui.adsabs.harvard.edu/abs/2019MNRAS.487.5610S} {487, 5610}

\bibitem[\protect\citeauthoryear{{Shahaf}, {Hallakoun}, {Mazeh}, {Ben-Ami},
  {Rekhi}, {El-Badry}  \& {Toonen}}{{Shahaf} et~al.}{2023}]{Shahaf_2023b}
{Shahaf} S.,  {Hallakoun} N.,  {Mazeh} T.,  {Ben-Ami} S.,  {Rekhi} P.,
  {El-Badry} K.,   {Toonen} S.,  2023, \mn@doi [arXiv e-prints]
  {10.48550/arXiv.2309.15143}, \href
  {https://ui.adsabs.harvard.edu/abs/2023arXiv230915143S} {p. arXiv:2309.15143}

\bibitem[\protect\citeauthoryear{{Skrutskie} et~al.,}{{Skrutskie}
  et~al.}{2006}]{2MASS}
{Skrutskie} M.~F.,  et~al., 2006, \mn@doi [\aj] {10.1086/498708}, \href
  {https://ui.adsabs.harvard.edu/abs/2006AJ....131.1163S} {131, 1163}

\bibitem[\protect\citeauthoryear{{Tian}, {El-Badry}, {Rix}  \& {Gould}}{{Tian}
  et~al.}{2020}]{Tian_2020}
{Tian} H.-J.,  {El-Badry} K.,  {Rix} H.-W.,   {Gould} A.,  2020, \mn@doi
  [\apjs] {10.3847/1538-4365/ab54c4}, \href
  {https://ui.adsabs.harvard.edu/abs/2020ApJS..246....4T} {246, 4}

\bibitem[\protect\citeauthoryear{Toonen, Hollands, Gänsicke  \&
  Boekholt}{Toonen et~al.}{2017}]{Toonen_2017}
Toonen S.,  Hollands M.,  Gänsicke B.~T.,   Boekholt T.,  2017, \mn@doi [A\&A]
  {10.1051/0004-6361/201629978}, 602, A16

\bibitem[\protect\citeauthoryear{Vos, Vučković, Chen, Han, Boudreaux, Barlow,
  Østensen  \& Németh}{Vos et~al.}{2018}]{Vos_2018}
Vos J.,  Vučković M.,  Chen X.,  Han Z.,  Boudreaux T.,  Barlow B.~N.,
  Østensen R.,   Németh P.,  2018, \mn@doi [Monthly Notices of the Royal
  Astronomical Society] {10.1093/mnras/sty3017}, 482, 4592

\bibitem[\protect\citeauthoryear{{Webbink}}{{Webbink}}{2008}]{Webbink_2008}
{Webbink} R.~F.,  2008, in {Milone} E.~F.,  {Leahy} D.~A.,   {Hobill} D.~W.,
  eds,  Astrophysics and Space Science Library Vol. 352, Astrophysics and Space
  Science Library. p.~233 (\mn@eprint {arXiv} {0704.0280}),
  \mn@doi{10.1007/978-1-4020-6544-6_13}

\bibitem[\protect\citeauthoryear{{Willems, B.} \& {Kolb, U.}}{{Willems, B.} \&
  {Kolb, U.}}{2004}]{Willems_2004}
{Willems, B.} {Kolb, U.} 2004, \mn@doi [A\&A] {10.1051/0004-6361:20040085},
  419, 1057

\bibitem[\protect\citeauthoryear{{Yamaguchi} et~al.,}{{Yamaguchi}
  et~al.}{2023}]{Yamaguchi_2023}
{Yamaguchi} N.,  et~al., 2023, \mn@doi [arXiv e-prints]
  {10.48550/arXiv.2309.15905}, \href
  {https://ui.adsabs.harvard.edu/abs/2023arXiv230915905Y} {p. arXiv:2309.15905}

\bibitem[\protect\citeauthoryear{{Zorotovic} \& {Schreiber}}{{Zorotovic} \&
  {Schreiber}}{2022}]{Zorotovic_2022}
{Zorotovic} M.,  {Schreiber} M.,  2022, \mn@doi [\mnras]
  {10.1093/mnras/stac1137}, \href
  {https://ui.adsabs.harvard.edu/abs/2022MNRAS.513.3587Z} {513, 3587}

\bibitem[\protect\citeauthoryear{{Zorotovic}, {Schreiber, M. R.}, {G\"ansicke,
  B. T.}  \& {Nebot G\'omez-Mor\'an, A.}}{{Zorotovic}
  et~al.}{2010}]{Zorotovic_2010}
{Zorotovic} M.,  {Schreiber, M. R.} {G\"ansicke, B. T.}  {Nebot
  G\'omez-Mor\'an, A.} 2010, \mn@doi [A\&A] {10.1051/0004-6361/200913658}, 520,
  A86

\bibitem[\protect\citeauthoryear{Zorotovic, Schreiber, Garc\'{\i}a-Berro,
  Camacho, Torres, Rebassa-Mansergas  \& G\"ansicke}{Zorotovic
  et~al.}{2014}]{Zorotovic_2014}
Zorotovic M.,  Schreiber M.~R.,  Garc\'{\i}a-Berro E.,  Camacho J.,  Torres S.,
   Rebassa-Mansergas A.,   G\"ansicke B.~T.,  2014, \mn@doi [A\&A]
  {10.1051/0004-6361/201323039}, 568, A68

\makeatother
\end{thebibliography}




\appendix

\section{Tables of Parameters}
\label{sec:tables_of_Params}

\begin{landscape}
\begin{table}
    \label{tab:Astro}
    \centering
    \caption{Table of astrometric binary systems, separated by their original survey. Here, \textit{RAVE}, \textit{LAMOST} and \textit{TGAS} are used as shorthand for \citet{PathwayI}, \citet{PathwayII} and \citet{PathwayV} respectively. The full table consisting of all 20 \textit{RAVE} astrometric binaries, 39 \textit{LAMOST} astrometric binaries and 64 \textit{TGAS} astrometric binaries, with a more complete list of properties, will be made available online upon publication.}
    \begin{tabular}{@{}lccccccccc@{}}
        \hline 

        Name & $P_\mathrm{orb}$ [d] & $T_\mathrm{eff, LS}$ [K] & log($g$)$_\mathrm{LS}$ [dex] & $R_\mathrm{LS}$ [R\textsubscript{$\odot$}] & $M_\mathrm{LS}$ [M\textsubscript{$\odot$}] & $T_\mathrm{eff, WD}$ [K] & $M_\mathrm{WD}$ [M\textsubscript{$\odot$}] & Contaminant & Original Survey \\
        \hline

        BD-13 6521 & 459.7 $\pm$ 2.0 & 6248.8 $\pm$ 37.8 & 4.424 $\pm$ 0.022 & 1.037 $\pm$ 0.016 & 1.039 $\pm$ 0.036 & 25525.6 $\pm$ 895.8 & 0.479 $\pm$ 0.024 & True & \textit{RAVE} \\
        
        TYC 4670-766-1 & 547.7 $\pm$ 0.598 & 5372.5 $\pm$ 29.8 & 4.409 $\pm$ 0.007 & 0.974 $\pm$ 0.011 & 0.888 $\pm$ 0.021 & 11018.0 $\pm$ 94.8 & 0.456 $\pm$ 0.009 & False & \textit{RAVE}\\
        
        TYC 5202-162-1 & 826.5 $\pm$ 61.9 & 5888.8 $\pm$ 43.5 & 3.498 $\pm$ 0.021 & 3.821 $\pm$ 0.112 & 1.674 $\pm$ 0.033 & 31222.2 $\pm$ 5208.8 & 0.421 $\pm$ 0.072 & False & \textit{RAVE} \\
        
        UCAC2 28312072 & 442.9 $\pm$ 0.5 & 5154.5 $\pm$ 14.0 & 4.593 $\pm$ 0.007 & 0.777 $\pm$ 0.004 & 0.865 $\pm$ 0.010 & 12014.4 $\pm$ 227.5 & 0.578 $\pm$ 0.005 & True & \textit{RAVE} \\
        
        TYC 6086-1317-1 & 524.2 $\pm$ 3.3 & 6763.8 $\pm$ 44.5 & 4.106 $\pm$ 0.017 & 1.679 $\pm$ 0.018 & 1.310 $\pm$ 0.035 & 20844.7 $\pm$ 378.6 & 0.377 $\pm$ 0.012 & False & \textit{RAVE} \\

        ... & ... & ... & ... & ... & ... & ... & ... & ... & ... \\

        \hline

        Lamost J111853.41-084457.3 & 718.0 $\pm$ 4.4 & 6070.9 $\pm$ 40.1 & 4.361 $\pm$ 0.015 & 1.234 $\pm$ 0.019 & 1.271 $\pm$ 0.023 & 12940.0 $\pm$ 231.2 & 0.492 $\pm$ 0.018 & True & \textit{LAMOST} \\
        
        Lamost J035311.09+290033.4 & 1484.3 $\pm$ 134.1 & 5172.5 $\pm$ 48.6 & 2.874 $\pm$ 0.041 & 8.467 $\pm$ 0.144 & 1.958 $\pm$ 0.141 & 52999.1 $\pm$ 10745.2 & 1.109 $\pm$ 0.162 & False & \textit{LAMOST} \\
        
        TYC 1749-1463-1 & 718.3 $\pm$ 3.2 & 6098.2 $\pm$ 50.2 & 4.139 $\pm$ 0.021 & 1.433 $\pm$ 0.017 & 1.030 $\pm$ 0.038 & 28958.9 $\pm$ 1506.8 & 0.561 $\pm$ 0.017 & False & \textit{LAMOST} \\
        
        Lamost J084109.22+254236.1 & 464.1 $\pm$ 1.7 & 5738.6 $\pm$ 27.3 & 4.494 $\pm$ 0.012 & 0.951 $\pm$ 0.008 & 1.028 $\pm$ 0.020 & 14748.5 $\pm$ 333.1 & 0.515 $\pm$ 0.010 & False & \textit{LAMOST} \\
        
        Lamost J083358.62+333406.8 & 530.6 $\pm$ 5.7 & 6780.1 $\pm$ 69.6 & 4.248 $\pm$ 0.020 & 1.400 $\pm$ 0.024 & 1.261 $\pm$ 0.030 & 26649.8 $\pm$ 1476.5 & 0.438 $\pm$ 0.033948 & False & \textit{LAMOST} \\

        ... & ... & ... & ... & ... & ... & ... & ... & ... & ... \\

        \hline

        TYC 6465-1734-1 & 1285.0 $\pm$ 45.1 & 6196.1 $\pm$ 29.1 & 4.419 $\pm$ 0.016 & 1.063 $\pm$ 0.007 & 1.081 $\pm$ 0.030 & 19320.6 $\pm$ 1955.3 & 0.547 $\pm$ 0.041 & False & \textit{TGAS} \\
        
        TYC 9148-665-1 & 1319.1 $\pm$ 56.4 & 5547.2 $\pm$ 23.5 & 4.568 $\pm$ 0.010 & 0.815 $\pm$ 0.004 & 0.894 $\pm$ 0.016 & 15323.5 $\pm$ 2813.9 & 0.596 $\pm$ 0.039 & False & \textit{TGAS} \\
        
        TYC 9344-137-1 & 164.2 $\pm$ 1.1 & 4751.0 $\pm$ 30.9 & 3.119 $\pm$ 0.038 & 4.626 $\pm$ 0.052 & 1.028 $\pm$ 0.085 & 5081.9 $\pm$ 526.8 & 0.184 $\pm$ 0.023 & True & \textit{TGAS} \\
        
        TYC 8823-1109-1 & 820.0 $\pm$ 8.1 & 6455.1 $\pm$ 41.0 & 4.253 $\pm$ 0.016 & 1.394 $\pm$ 0.013 & 1.265 $\pm$ 0.029 & 29190.5 $\pm$ 1464.8 & 0.528 $\pm$ 0.010 & False & \textit{TGAS} \\
        
        TYC 8384-1121-1 & 793.4 $\pm$ 12.0 & 6680.4 $\pm$ 75.3 & 4.200 $\pm$ 0.019 & 1.506 $\pm$ 0.025 & 1.308 $\pm$ 0.030 & 24888.5 $\pm$ 901.7 & 0.496 $\pm$ 0.019 & False & \textit{TGAS} \\

        ... & ... & ... & ... & ... & ... & ... & ... & ... & ...\\

        \hline
    \end{tabular}
\end{table}
\end{landscape}

\begin{landscape}
\begin{table}
    \label{tab:Spectro}
    \centering
    \caption{Table of spectroscopic binary systems, separated by their original survey. Here, \textit{RAVE}, \textit{LAMOST} and \textit{TGAS} are used as shorthand for \citet{PathwayI}, \citet{PathwayII} and \citet{PathwayV} respectively. The full table consisting of all 35 \textit{RAVE} spectroscopic binaries, 32 \textit{LAMOST} spectroscopic binaries and 64 \textit{TGAS} spectroscopic binaries, with a more complete list of properties, will be made available online upon publication.}
    \begin{tabular}{@{}lccccccccc@{}}
        \hline 

        Name & $P_\mathrm{orb}$ [d] & $T_\mathrm{eff, LS}$ [K] & log($g$)$_\mathrm{LS}$ [dex] & $R_\mathrm{LS}$ [R\textsubscript{$\odot$}] & $M_\mathrm{LS}$ [M\textsubscript{$\odot$}] & $T_\mathrm{eff, WD}$ [K] & $M_\mathrm{WD}$ [M\textsubscript{$\odot$}] & Contaminant & Original Survey\\
        \hline

        TYC 6992-827-1 & 41.3 $\pm$ 0.1 & 5249.1 $\pm$ 28.1 & 3.520 $\pm$ 0.027 & 3.615 $\pm$ 0.113 & 1.578 $\pm$ 0.086 & - & 0.128 $\pm$ 0.010 & False & \textit{RAVE} \\
        
        TYC 6419-603-1 & 37.5 $\pm$ 0.0 & 4409.2 $\pm$ 13.9 & 2.647 $\pm$ 0.016 & 8.014 $\pm$ 0.124 & 1.039 $\pm$ 0.032 & 5800.6 $\pm$ 780.1 & 0.228 $\pm$ 0.020 & False & \textit{RAVE} \\
        
        TYC 5263-340-1 & 5.7 $\pm$ 0.0 & 4813.7 $\pm$ 14.0 & 4.527 $\pm$ 0.010 & 0.823 $\pm$ 0.004 & 0.833 $\pm$ 0.015 & 5526.2 $\pm$ 170.4 & 0.227 $\pm$ 0.004 & False & \textit{RAVE} \\
        
        BD-16 210 & 19.5 $\pm$ 0.0 & 4541.0 $\pm$ 9.9 & 2.669 $\pm$ 0.017 & 7.557 $\pm$ 0.108 & 0.972 $\pm$ 0.024 & - & 0.132 $\pm$ 0.004 & False & \textit{RAVE} \\
        
        UCAC3 161-284029 & 2.1 $\pm$ 0.0 & 4059.4 $\pm$ 16.6 & 3.812 $\pm$ 0.063 & 1.088 $\pm$ 0.008 & 0.283 $\pm$ 0.036 & - & 0.054 $\pm$ 0.005 & True & \textit{RAVE} \\

        ... & ... & ... & ... & ... & ... & ... & ... & ... & ... \\

        \hline

        Lamost J043826.34+394739.7 & 5.1 $\pm$ 0.0 & 5457.9 $\pm$ 74.9 & 2.138 $\pm$ 0.053 & 25.040 $\pm$ 1.935 & 3.112 $\pm$ 0.112 & 5908.7 $\pm$ 2260.8 & 0.150 $\pm$ 0.031 & False & \textit{LAMOST} \\
        
        Lamost J053056.93+460345.7 & 60.1 $\pm$ 0.1 & 5441.3 $\pm$ 214.3 & 2.665 $\pm$ 0.154 & 10.600 $\pm$ 0.345 & 1.983 $\pm$ 0.556 & - & 1.818 $\pm$ 0.281 & True & \textit{LAMOST} \\
        
        Lamost J054307.98+502435.6 & 45.2 $\pm$ 0.2 & 6019.7 $\pm$ 48.0 & 2.293 $\pm$ 0.087 & 11.482 $\pm$ 0.395 & 0.884 $\pm$ 0.205 & 29063.6 $\pm$ 12979.4 & 0.318 $\pm$ 0.056 & False & \textit{LAMOST} \\
        
        Lamost J051005.99+492321.9 & 146.3 $\pm$ 0.5 & 5436.0 $\pm$ 76.9 & 2.046 $\pm$ 0.055 & 25.608 $\pm$ 1.801 & 2.651 $\pm$ 0.184 & - & 0.954 $\pm$ 0.050 & False & \textit{LAMOST} \\
        
        Lamost J050413.97+530343.8 & 3.2 $\pm$ 0.0 & 5215.0 $\pm$ 137.4 & 2.495 $\pm$ 0.108 & 11.344 $\pm$ 0.460 & 1.515 $\pm$ 0.403 & 5903.1 $\pm$ 2285.2 & 0.164 $\pm$ 0.036 & False & \textit{LAMOST} \\

        ... & ... & ... & ... & ... & ... & ... & ... & ... & ...\\

        \hline

        TYC 2562-1312-1 & 16.3 $\pm$ 0.0 & 6084.4 $\pm$ 40.7 & 4.418 $\pm$ 0.021 & 1.030 $\pm$ 0.009 & 1.012 $\pm$ 0.036 & - & 0.172 $\pm$ 0.008 & True & \textit{TGAS} \\
        
        TYC 1241-181-1 & 1.7 $\pm$ 0.0 & 6397.8 $\pm$ 79.3 & 3.634 $\pm$ 0.028 & 2.314 $\pm$ 0.026 & 0.867 $\pm$ 0.035 & 19321.1 $\pm$ 810.8 & 0.456 $\pm$ 0.020 & False & \textit{TGAS} \\
        
        TYC 1218-1456-1 & 1.3 $\pm$ 0.0 & 7729.5 $\pm$ 56.0 & 3.995 $\pm$ 0.029 & 2.133 $\pm$ 0.028 & 1.639 $\pm$ 0.092 & - & 0.167 $\pm$ 0.009 & False & \textit{TGAS} \\
        
        TYC 1783-665-1 & 18.0 $\pm$ 0.0 & 5708.7 $\pm$ 14.2 & 4.538 $\pm$ 0.010 & 0.818 $\pm$ 0.012 & 0.845 $\pm$ 0.008 & - & 0.130 $\pm$ 0.011 & False & \textit{TGAS} \\
        
        TYC 2803-978-1 & 1035.9 $\pm$ 218.5 & 6210.1 $\pm$ 40.1 & 4.118 $\pm$ 0.038 & 1.589 $\pm$ 0.078 & 1.203 $\pm$ 0.049 & 10159.4 $\pm$ 3025.5 & 0.291 $\pm$ 0.049 & True & \textit{TGAS} \\

        ... & ... & ... & ... & ... & ... & ... & ... & ... & ...\\
        
        \hline
    \end{tabular}
\end{table}


\bsp	
\label{lastpage}
\end{landscape}
\end{document}